\UseRawInputEncoding
\RequirePackage{rotating}
\documentclass[fleqn,usenatbib]{mnras}
\usepackage{epsfig}
\usepackage{amsmath}
\usepackage{graphicx}
\usepackage{array}
\usepackage{textcomp}
\usepackage{amssymb}
\usepackage{rotating}
\usepackage{pdflscape}
\usepackage{caption}
\usepackage{subcaption}
\usepackage{longtable}
\usepackage{xtab}
\usepackage{xcolor}
\usepackage{colortbl}
\usepackage{hhline}
\def\apjs {ApJS}
\def\apjl {ApJL}

\def\aap {A\&A}

\DeclareRobustCommand{\ion}[2]{%
\relax\ifmmode
\ifx\testbx\f@series
{\mathbf{#1\,\mathsc{#2}}}\else
{\mathrm{#1\,\mathsc{#2}}}\fi
\else\textup{#1\,{\mdseries\textsc{#2}}}%
\fi}

\title[Gas and stellar metallicities on the SFMS]
      {The SAMI Galaxy Survey: the drivers of gas and stellar metallicity differences in galaxies.}  
\author[A.\ Fraser-McKelvie et al.]
       {A.\ Fraser-McKelvie$^{1,2}$\thanks{a.fraser-mckelvie@uwa.edu.au}, L.\ Cortese$^{1,2}$, B.\ Groves$^{1,2}$, S.\ Brough$^{3,2}$, J.\ Bryant$^{4,5,2}$,  B.\ Catinella$^{1,2}$, \and S.\ Croom$^{4,2}$, F.\ D'Eugenio$^{6,7}$, \'{A}.\ R.\ L\'{o}pez-S\'{a}nchez$^{8,9,2}$, J.\ van de Sande$^{4,2}$, S.\ Sweet$^{10,2}$, S.\ Vaughan$^{4}$, \and J.\ Bland-Hawthorn$^{4}$, J.\ Lawrence$^{8}$, N.\ Lorente$^{11}$, and M.\ Owers$^{9,12,2}$.       
                \vspace*{1mm}\\
        $^{1}$ International Centre for Radio Astronomy Research, The University of Western Australia, 35 Stirling Hwy, 6009 Crawley, WA, Australia \\
        $^{2}$ ARC Centre of Excellence for All Sky Astrophysics in 3 Dimensions (ASTRO 3D) \\
        $^{3}$ School of Physics, University of New South Wales, NSW 2052, Australia \\ 
        $^{4}$ Sydney Institute for Astronomy (SIfA), School of Physics, The University of Sydney, NSW 2006, Australia \\
        $^{5}$ Australian Astronomical Optics, AAO-USydney, School of Physics, University of Sydney, NSW 2006, Australia \\
        $^{6}$ Cavendish Laboratory and Kavli Institute for Cosmology, University of Cambridge, Madingley Rise, Cambridge, CB3 0HA, United Kingdom \\
        $^{7}$ Sterrenkundig Observatorium, Universiteit Gent, Krijgslaan 281 S9, B-9000 Gent, Belgium \\
        $^{8}$ Australian Astronomical Optics, Macquarie University, 105 Delhi Rd, North Ryde, NSW 2113, Australia \\ 
        $^{9}$ Department of Physics and Astronomy, Macquarie University, NSW 2109, Australia \\
        $^{10}$ School of Mathematics and Physics, University of Queensland, Brisbane, QLD 4072, Australia \\
        $^{11}$ AAO-MQ, Faculty of Science \& Engineering, Macquarie University. 105 Delhi Rd, North Ryde, NSW 2113, Australia \\
        $^{12}$ Astronomy, Astrophysics and Astrophotonics Research Centre, Macquarie University, Sydney, NSW 2109, Australia \\
	}
\begin{document}
\maketitle
\begin{abstract}
The combination of gas-phase oxygen abundances and stellar metallicities can provide us with unique insights into the metal enrichment histories of galaxies. 
In this work, we compare the stellar and gas-phase metallicities measured within a 1$R_{e}$ aperture for a representative sample of 472 star-forming galaxies extracted from the SAMI Galaxy Survey. We confirm that the stellar and interstellar medium (ISM) metallicities are strongly correlated, with scatter $\sim$3 times smaller than that found in previous works, and that integrated stellar populations are generally more metal-poor than the ISM, especially in low-mass galaxies. The ratio between the two metallicities strongly correlates with several integrated galaxy properties including stellar mass, specific star formation rate, and a gravitational potential proxy. However, we show that these trends are primarily a consequence of: (a) the different star formation and metal enrichment histories of the galaxies, and (b) the fact that while stellar metallicities trace primarily iron enrichment, gas-phase metallicity indicators are calibrated to the enrichment of oxygen in the ISM. Indeed, once both metallicities are converted to the same `element base' all of our trends become significantly weaker. Interestingly, the ratio of gas to stellar metallicity is always below the value expected for a simple closed-box model, which requires that outflows and inflows play an important role in the enrichment history across our entire stellar mass range. 
This work highlights the complex interplay between stellar and gas-phase metallicities and shows how care must be taken in comparing them to constrain models of galaxy formation and evolution.

\end{abstract}


\section{Introduction}
The evolution of metal content in galaxies follows a broadly predictable route: every generation of stars is created from the \mbox{interstellar} medium (ISM), produces metals, and then expels these metals back into the ISM through supernovae or stellar winds. The next generation of stars is created from the \mbox{enriched} ISM, and they themselves will be more enriched than the previous stellar generation. Gas inflows and outflows complicate this otherwise `closed box' scenario by providing supplies of pristine gas and ejecting metal-enriched gas into the inter-galactic medium.

For a closed-box model of galactic chemical evolution we can assume that a galaxy increases its metal content over time. The gas-phase and stellar metallicities can then be calculated from a cumulative metallicity distribution: the gas-phase metallicity will be the maximum value of the distribution, and the stellar metallicity the (light- or mass-weighted) mean. The maximum will always be greater than the mean value, and hence the average stellar metallicity for a given galaxy is lower than that of the gas-phase metallicity when a closed-box model is assumed. One might imagine that the shape of the cumulative metallicity distribution will dictate the difference between the maximum and mean, and hence the difference between stellar and gas-phase metallicity values for a given galaxy. 

Observationally, measuring the metal content of a galaxy's constituents is not trivial, however it is a worthwhile exercise. Galaxy spectra are the weighted summation of the light from all generations of stars within a region sampled, and so the stellar metallicity is something of an average of the metal content of all of the stars within a galaxy.
The ionised gas emission lines we observe from star-forming regions are generally powered by young, hot stars, and as such reflect the properties of the galaxy at $z=0$. Combining gas and stellar metallicity measurements is therefore a powerful tool to constrain the evolutionary pathways of galaxies throughout all of cosmic time.

The metallicity of stars and gas are measured in a different manner to one another for several reasons. While oxygen is the most abundant element in the Universe (after hydrogen and helium), the abundance of this element in stars is difficult to measure. First, oxygen does not give rise to strong observable features in the optical stellar continuum spectra, and the absorption lines we do observe are either blended with other elemental lines \citep[even at high spectral resolution,][]{asplund2005}, or arise outside of local thermodynamical equilibrium \citep[which is difficult to model,][]{asplund2009}. Moreover, oxygen in cool stellar atmospheres depletes onto molecules \citep[e.g. OH, CO, TiO,][]{asplund2005}, which further complicates inferring the total abundance. For these reasons, estimating oxygen abundances in stars is problematic. Despite recent developments in stellar astrophysics \citep[e.g.][]{ting2018}, estimating the metallicity of whole stellar populations requires the further step of averaging over a distribution of stellar spectra which may have different abundances. For this reason, measuring galaxy stellar metallicities relies on the much more common and stronger iron absorption lines \citep{worthey1994}.
While iron is a reliable abundance indicator for stars, it cannot be used to determine the metal content of the ISM as it depletes readily onto dust \citep[e.g.][]{jenkins2009,nicholls2017}.
Instead, we define the gas metallicity as the oxygen abundance, which can be estimated from strong emission lines \citep[e.g.][]{searle1971, lequeux1979, pagel1979}. 
Employing different tracer elements means that stellar and gas metallicities are measured using different abundance scales, making comparisons difficult \citep[e.g.][]{peng2014}.

When seeking to understand the chemical evolutionary history of galaxies, many previous works have studied the link between either the gas-phase oxygen abundance or stellar metallicity with other galaxy observables. 
For example, there are a plethora of previous literature on the origin of the stellar mass--gas-phase oxygen abundance relation \citep[$M_{\star}-12+\log(\rm{O}/\rm{H})$ relation, often referred to as the $M_{\star}-Z_{gas}$ relation in the literature; e.g.][and references within]{lequeux1979,skillman1989, tremonti2004a, kewley2008, maiolino2019} and the stellar mass--stellar metallicity relation \citep[$M_{\star}-Z_{\star}$ relation; e.g.][]{faber1973, brodie1991, trager2000, gallazzi2005, zahid2017, lian2018}.
A positive correlation is observed in both gas and stars such that more massive (or luminous) galaxies possess a greater fraction of metals than their lower-mass counterparts.
The $M_{\star}-Z_{gas}$ relation bends at high stellar masses \citep[e.g.][]{tremonti2004a}. Rather than a saturation of the chosen gas-phase metallicity indicator, this flattening is thought to be due to a characteristic mass (corresponding to a characteristic gravitational potential well depth), above which galaxies are able to retain a large fraction of their metals within their halo. At lower stellar masses, feedback becomes effective at ejecting metals from the ISM via outflows \citep[e.g.][]{blanc2019}.
The source of these outflows is thought to be supernova explosions \citep[e.g.][]{larson1974,kobayashi2007, scannapieco2008}.
In support of this hypothesis, observationally we see that in addition to the stellar mass dependence, both gas-phase oxygen abundances and the stellar metallicities of a galaxy are strongly correlated with gravitational potential and local stellar surface mass density proxies \citep[e.g.][]{scott2017, barone2018, deugenio2018, gao2018, barone2020}.

Given the difference in timescales that stellar and gas metallicity probe, it is essential to include both in studies to understand the complete metallicity evolution of a galaxy. In addition, only through the inclusion of both metallicities can theoretical models of galaxy evolution be constrained.
Some early Sloan Digital Sky Survey \citep{abazajian2004} work investigated the differences between stellar metallicities and gas-phase oxygen abundances within the same galaxies \citep{gallazzi2005}. This work found that the stellar metallicity and gas-phase oxygen abundance of a galaxy are correlated. The substantial scatter in stellar metallicity at fixed oxygen abundance was explained by gas inflows/outflows within an individual galaxy. 
Indeed, analytic models require inflows and outflows to reliably model the cosmic metallicity evolution in galaxies \citep[e.g.][]{finlator2008, lilly2013, peng2014}.

In general, the average stellar metallicity for a given galaxy is found to be lower than its gas-phase oxygen abundance, with
the largest differences seen at low stellar masses \citep[e.g.][]{gonzalezdelgado2014, lian2018}. Indeed, the metallicity of the youngest stars most closely matches that of the gas-phase. \citet{lian2018} explain the large differences between stellar metallicity and gas-phase oxygen abundances at low stellar masses by expecting significant metallicity evolution throughout the life of a galaxy. To correctly model the observed differences between the metal content of stars and gas they invoke models with either strong metal-rich outflows, or steep initial mass functions (IMF), both of which are quite extreme scenarios. 

The large difference between stellar and gas-phase metallicities at low stellar masses and the reason why the gap between the two closes at higher stellar masses are both open questions. One further possible explanation could be a variety of star formation histories (and therefore chemical enrichment histories) in individual galaxies.
Previous work has explored the fossil record of the $M_{\star}-Z_{\star}$ relation via full spectral fitting \citep[e.g.][]{fernandes2007,panter2008,valeasari2009}. These works found the $M_{\star}-Z_{\star}$ relation to be steeper in older stellar populations, and that more massive galaxies show very little evolution since distant epochs \citep[however this is not the case when observing star-forming vs. passive galaxies in the current epoch e.g.][]{peng2015}. The rapid early growth of high-mass galaxies (referred to as a `chemical downsizing' scenario) can be used to explain the differences in the stellar metallicity of old and young stellar populations today. 
That said, these works rely on full spectral fitting providing reliable metallicity estimates for young stellar populations and provide no constraints on the gas itself.

Another key point generally overlooked by previous works is that gas and stellar metallicity estimates are biased towards elements that form over different timescales.
As mentioned already, ionised gas abundances are usually calibrated to oxygen, which is an $\alpha$ element, while iron, a non-$\alpha$ element, is often used to infer stellar metallicities. 
$\alpha$ elements are heavy elements mostly produced in massive stars and ejected by core-collapse supernovae on short timescales of $\sim$10 Myr \citep[e.g.][]{timmes1995}. Non-$\alpha$ elements such as iron-peak elements are formed by Type Ia supernovae over longer timescales \citep[e.g.][]{kobayashi2009}. Hence, the ratio of $\alpha$ to non-$\alpha$ elements will vary as a function of time for a galaxy depending on its star formation history (SFH).
This discrepancy needs to be taken into account when comparing gas and stellar metallicities and as such, blindly equating stellar metallicity to an oxygen abundance does not really provide a fair comparison between the enrichment of gas and stars.

We aim to build on previous work and combine stellar and gas-phase metallicity information with the intention of determining the driving mechanism behind any differences seen. 
In addition, we wish to gain a feel for the effects of a galaxy's SFH on the trends we observe
by converting oxygen abundances (which we term $Z_{gas}$ in line with previous literature) to a total gas metallicity (which we term $Z_{gas,tot}$) to compare to the stellar metallicity.
For this work we require a large sample of galaxies with both reliable stellar population and gas-phase oxygen abundance indicators. Such measurements are only available for star-forming galaxies with excellent continuum signal-to-noise.
In addition, to build on early fibre spectroscopy work, it would be advantageous to examine integrated quantities of  galaxies, to understand the evolution of metal content across entire galaxies and not just their nuclear regions. The Sydney-AAO Multi-object Integral field spectrograph \citep[SAMI;][]{croom2012} Galaxy Survey fits this brief perfectly. 

In this paper we describe the SAMI Galaxy Survey and the gas and stellar metallicity indicators used in this work in Section~\ref{data}, compare the two and seek to understand the drivers behind the correlations seen in Section~\ref{results}, before discussing the observables driving the differences in the metal content of gas and stars in Section~\ref{discussion}. We summarise and conclude our results in Section~\ref{conclusion}. Throughout, we use a \citet{chabrier2003} IMF and $\Lambda$CDM cosmology, with $\Omega_{m}=0.3$, $\Omega_{\lambda}=0.7$, $H_{0}=70~\rm{km}~\rm{s}^{-1}~\rm{Mpc}^{-1}$.

\section{Data and methods}
\label{data}
\subsection{The SAMI Galaxy Survey}
The SAMI Galaxy Survey is an integral field spectroscopy survey on the Anglo-Australian Telescope that observed 3068 galaxies from 2013--2018 \citep{croom2012}. 
SAMI used 13 fused fibre hexabundles \citep{bland-hawthorn2011,bryant2014} with a high (75\%) fill factor. Each bundle contained 61 fibres of $1.6^{\prime\prime}$ diameter resulting in each integral field unit (IFU) having a diameter of $15^{\prime\prime}$. The IFUs, as well as 26 sky fibres, were plugged into pre-drilled plates using magnetic connectors. SAMI fibres were fed to the double-beam AAOmega spectrograph \citep{sharp2015}, which allowed a range of different resolutions and wavelength ranges. 
The SAMI Galaxy survey employed the 580V grating between 3750--5750\,\AA\ giving a resolution of R=1810 ($\sigma = 70.4~\rm{km}~\rm{s}^{-1}$) at 4800\,\AA\, and the 1000R grating from 6300--7400\,\AA\ giving a resolution of R=4260 ($\sigma = 29.6~\rm{km}~\rm{s}^{-1}$) at 6850\,\AA\ \citep{scott2018}. 83\% of galaxies in the SAMI target catalogue had coverage out to 1 effective radius \citep[1$R_{e}$,][]{bryant2015}.

The SAMI Galaxy Survey was comprised of a sample drawn from the Galaxy and Mass Assembly \citep[GAMA;][]{driver2011} survey equatorial regions \citep{bryant2015}, and an additional sample of eight clusters \citep{owers2017}. SAMI Data Release 3 \citep[DR3;][]{croom2021} contains observations of 3068 galaxies and is the final data release of the SAMI survey. 
SAMI DR3 includes observations spanning $0.04<z<0.128$ and $7.42<\log M_{\star}[\rm{M}_{\odot}] < 11.89$ (corresponding to an $r$-band magnitude range of $18.4 < m_{r} < 12.1$), with environments ranging from underdense field regions to extremely overdense clusters. 

The data products used in this analysis are the spectral cubes for full spectral fitting and aperture emission line maps for gas-phase oxygen abundance estimates.

 \begin{figure}
\centering
\begin{subfigure}{0.49\textwidth}
\includegraphics[width=\textwidth]{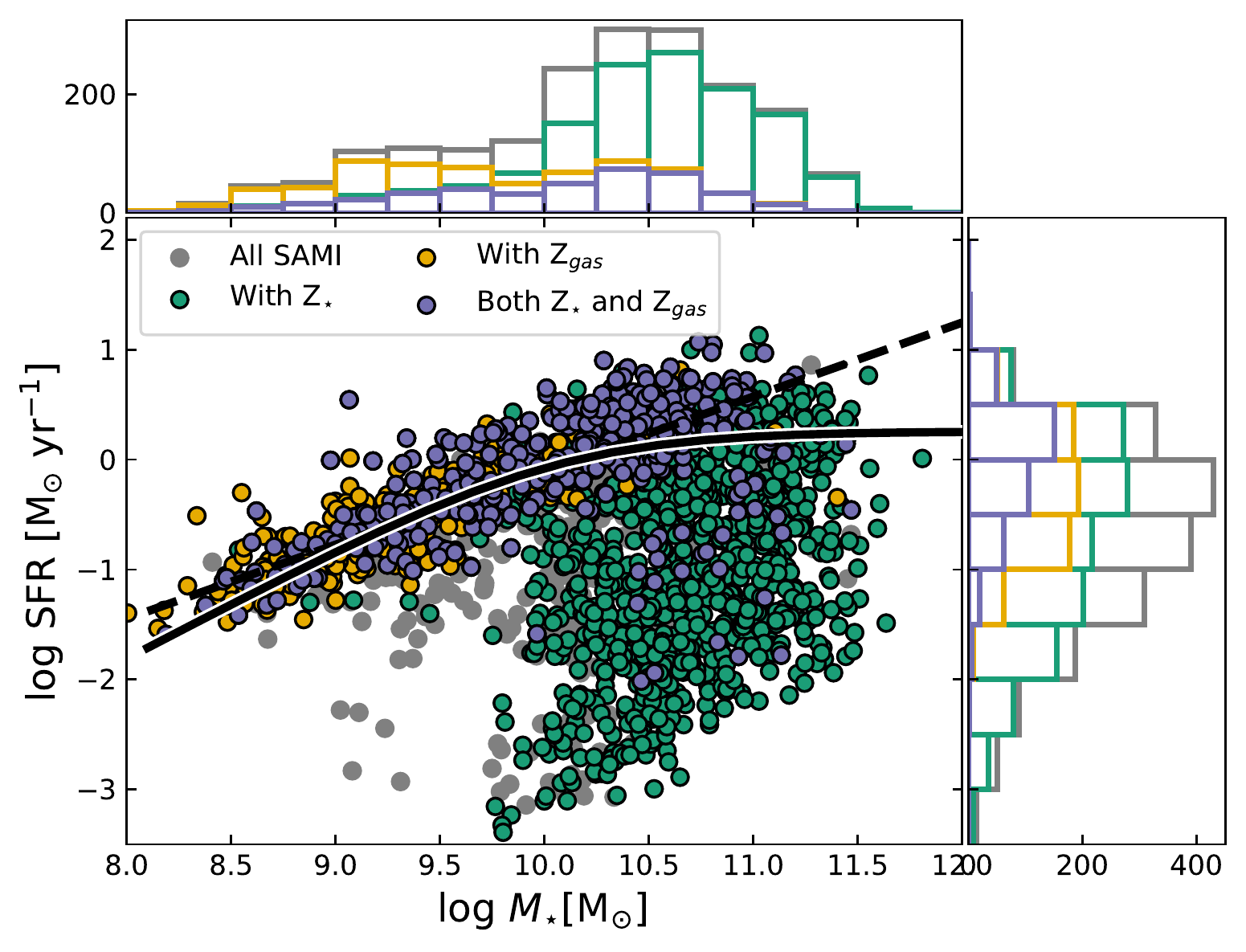}  
\end{subfigure}
\caption{The SAMI galaxies used in this work in the SFR-stellar mass plane. Histograms in mass and SFR for each population are shown to the top and right side of the plot, respectively. Purple points are 472 galaxies with $Z_{gas}$ and $Z_{\star}$ information, satisfy the S/N cuts discussed in the text and are the sample used in the following analysis. Orange points have only $Z_{gas}$, green points have only $Z_{\star}$, and grey points have neither indicator. The curved and linear star-forming main sequence lines of \citet{fraser-mckelvie_sami_2021} are shown in black for reference.}
\label{sample}
\end{figure}

\subsection{Star formation rates and stellar masses}
Star formation rates (SFRs) and stellar masses were taken from the $GALEX$-Sloan-$WISE$ Legacy Catalogue-2 \citep[GSWLC-2;][]{salim2016, salim2018} using a sky match with maximum allowed separation of 2$^{\prime\prime}$. In that catalogue, Stellar masses and SFRs were determined via SED fitting using the Code Investigating GALaxy Emission \citep[CIGALE;][]{noll2009, boquien2019}. The photometry included in the fits was sourced from the Wide-Field Survey Explorer ($WISE$) in the mid-infrared, Sloan Digital Sky Survey in the optical, and the Galaxy Evolution Explorer ($GALEX$) in the UV. 
For this work we employ the GSWLC-X2 catalogue which uses the deepest $GALEX$ photometry available for a given source (selected from the shallow `all-sky', medium-deep, and deep catalogues). 
As SDSS coverage is limited to declinations $\gtrsim -10$\textdegree, we do not have SFRs and stellar masses from the GSWLC-2 for the four Southern SAMI clusters. Despite this limitation, 1832 SAMI galaxies have SFRs and stellar masses from GSWLC-2.

 Figure~\ref{sample} shows the distribution of all SAMI galaxies with GSWLC-2 data in the SFR--$M{\star}$ plane (grey points). Galaxies with available stellar metallicity estimates (and satisfy the quality cuts detailed in Section~\ref{stellar_mets_sect}) are shown in green, and those with gas-phase oxygen abundances (and satisfy the quality cuts detailed in Section~\ref{gas_metallicities}) are shown in orange. Galaxies with both stellar and gas-phase abundance estimates are shown in purple. Unsurprisingly due to their need for both good continuum and emission line S/N, the majority of these galaxies are located on or around the star-forming main sequence lines shown in black. The purple points in Figure~\ref{sample} constitute the sample used for the analysis in this work.

\subsection{Stellar metallicity estimates}
\label{stellar_mets_sect}
Full spectral fitting was employed to determine light-weighted average ages and metallicities of the SAMI galaxies from the 1$R_{e}$ aperture spectra (for mass-weighted results, see Appendix~\ref{appendixa}). The integrated spectra were obtained by co-adding all spectra of spaxels within an elliptical aperture of 1$R_{e}$ semi-major axis, where the aperture was defined by the axial ratio determined from the multi-Gaussian expansion fits of \citet{deugenio2021} as in \citet{scott2018} and using the method of \citet{emsellem1994} and \citet{cappellari2002}.
We employ the penalised pixel-fitting (\textsc{pPXF}) method of \citet{cappellari2004} and \citet{cappellari2017}. We follow one of the examples included in the \textsc{pPXF} package closely (\texttt{ppxf\_example\_population.py}) after modifying it for use with SAMI data.

We adopted the method of \citet{vandesande2017} to prepare the SAMI 1$R_{e}$ aperture spectra for use with full spectral fitting codes. The spectra from the blue and red arms of the SAMI spectrograph were initially spliced after first convolving the red (higher resolution, $\sigma=29.6~ \rm{km}~\rm{s}^{-1}$) spectrum to the instrumental resolution of the blue arm ($\sigma = 70.4~ \rm{km}~\rm{s}^{-1}$). The resultant combined spectrum was then log rebinned. 

Emission lines and stellar continuum were fit simultaneously by creating a set of Gaussian emission line templates to fit at the same time as the stellar templates. In this manner, only the spectral gap between the SAMI red and blue arm wavelength coverage regions (5760--6400\,\AA) was masked. Barring the masked middle region, the spectrum was fit in the rest wavelength range 3900--6800\,\AA.
Regularisation was employed for the \textsc{pPXF} fit and the template normalisation for light-weighted properties performed over the range 5000--5500\,\AA. The `clean' keyword was set so that an iterative sigma clipping method was used to remove unmasked bad pixels from the spectra.

We employed the MILES SSP model templates \citep{vazdekis2010} with ages 0.08, 0.10, 0.13, 0.16, 0.20, 0.25, 0.32, 0.40, 0.50, 0.63, 0.79, 1.0, 1.26, 1.58, 2.0, 2.51, 3.16, 3.98, 5.01, 6.31, 7.94, 10.0, 12.59, 15.85 Gyr , metallicities -1.71, -1.31, -0.71, -0.4, 0.0, 0.22 , and `baseFe' (Milky Way value) $\alpha$ enhancement. 

 \begin{figure}
\centering
\begin{subfigure}{0.49\textwidth}
\includegraphics[width=\textwidth]{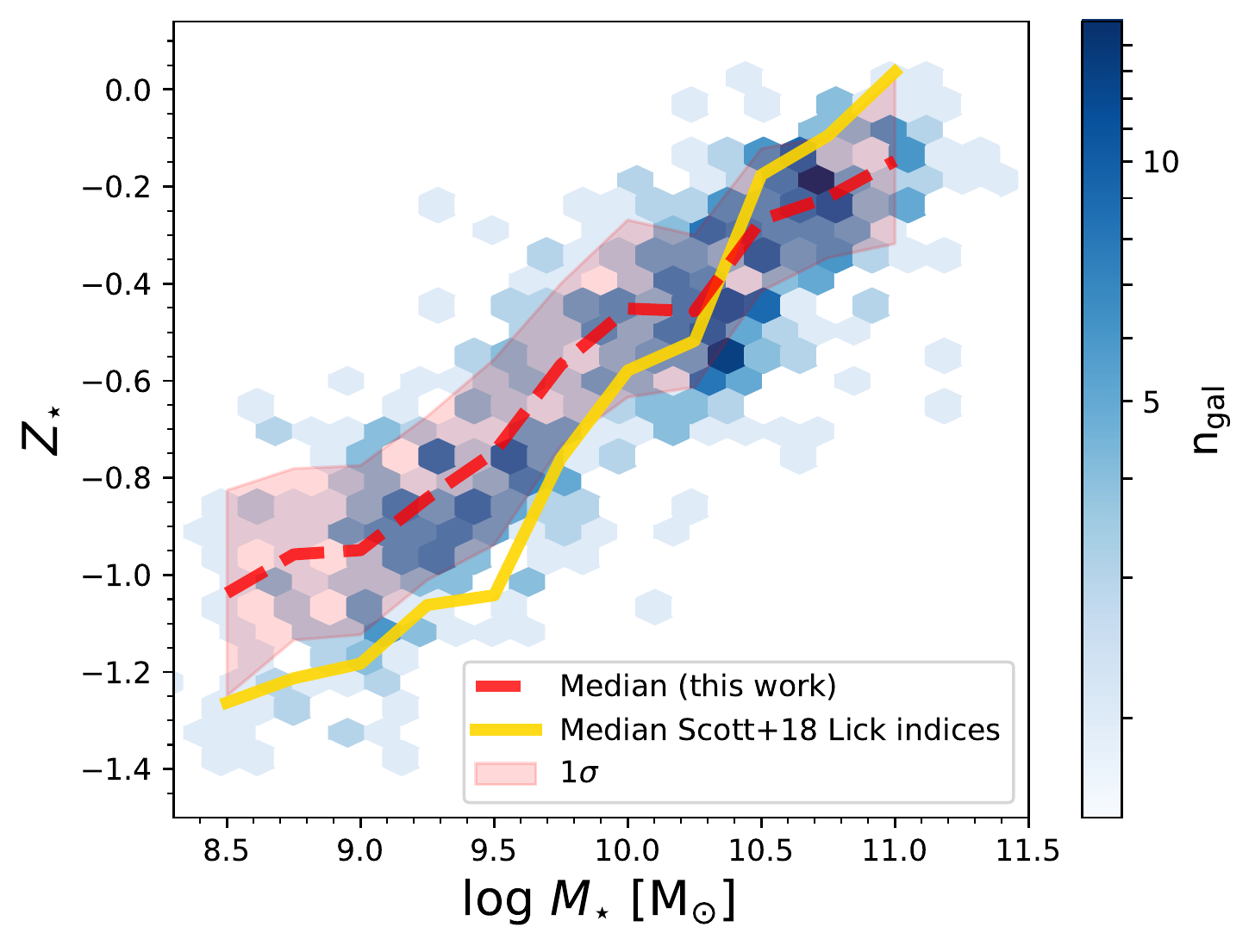}  
\end{subfigure}
\caption{Stellar mass-stellar metallicity relation ($M_{\star}-Z_{\star}$ relation) derived via light-weighted stellar population analysis using the full spectral fitting code \textsc{pPXF}. The mass-metallicity relation is apparent, with the median of the relation shown as a dashed red line and the shaded region is the 1$\sigma$ scatter. The average scatter in the $M_{\star}-Z_{\star}$ relation is 0.15 dex. For comparison, the median $M_{\star}-Z_{\star}$ relation for the same SAMI galaxies using the Lick index-derived stellar metallicities presented in \citet{scott2018} is shown in gold. }
\label{stell_mass_met}
\end{figure}

Previous work has shown that full spectrum fitting often struggles to fit an excess of blue light within a galaxy \citep[e.g.][]{cidfernandes2010} normally attributed to horizontal branch stars in the planetary nebula phase \citep[e.g.][]{yi2008}. As noted in \citet{peterken2020}, this light is not accounted for in old SSP template spectra, and so full spectral fitting software often attributes it to young, metal-poor populations. For our stellar population age and metallicity estimates, we therefore follow the method of \citet{peterken2020} and whilst we use the entire template range in our fits, we remove the contribution of the youngest SSP templates (in our case 0.08 Gyr) weights when calculating average ages and metallicities. \citet{peterken2020} show that older stellar populations are unaffected by the method by which the younger populations are modelled, and so are robust.

Given that the SSP templates are approximately evenly spaced in log age and metallicity space, and coupled with the large dynamic range between the oldest and youngest, and most metal-rich and metal-poor templates, we chose to compute the average metallicity via a logarithmic mean. Logarithmically-derived means are the default option within \textsc{pPXF}, and regularly utilised in previous literature \citep[e.g.][]{gonzalez-delgado2015}. We are most interested in metallicity measures for this work, and note that there will be an offset between the logarithmically- and linearly-derived mean metallicity measures for a given galaxy such that the mean logarithmically-derived metallicity value will be more metal-poor than that derived linearly.

The logarithmic mean of the age and metallicity per spectrum were computed from the output template weights following equations (1) and (2) of \citet{mcdermid2015} but excluding the youngest SSP models. The resultant [M/H] metallicity estimate we obtain is termed $Z_{\star}$, such that 
\begin{equation*}
    Z_{\star} \equiv [\rm{M/H}] = \log_{10}[(\rm{M/H})/(\rm{M/H})_{\odot}].
\end{equation*} 
A $\chi^{2}$ cut was employed such that galaxies with \textsc{pPXF} fits with reduced $\chi^{2}>1.5$ are not included in our results. This cut reduced the number of galaxies with stellar metallicity measures from 2861 to 2745, and the majority of those cut were either low-mass or low continuum signal to noise. 
We present the results of the light-weighted fits here as they are more directly comparable to observations and Lick index-derived results. 

We show the $M_{\star}-Z_{\star}$ relation for all galaxies in the SAMI sample with both $Z_{\star}$ and $Z_{gas}$ estimates in Figure~\ref{stell_mass_met}. Here the hexbins are coloured by the number of galaxies per bin. The median $Z_{\star}$ in bins of $M_{\star}$ is shown as the red dashed line with the 1$\sigma$ uncertainty shaded in red. A tight mass-metallicity relation is apparent, with a median $1\sigma$ scatter of 0.15 dex. 
As a comparison, we also extracted the 1$R_{e}$ aperture stellar population parameters from \citet{scott2017}, which were derived via Lick indices. The median $Z_{\star}$ obtained per mass bin from the Lick index measurements is shown as a gold line in Figure~\ref{stell_mass_met}. The mass-metallicity relation derived via Lick indices is steeper than the light-weighted \textsc{pPXF} relation, probably the result of both a larger range of metallicity SSP templates used in the \citet{scott2017} work, and the fact that Lick indices assume a single-burst star formation history. Non-parametric full spectral fitting tools such as \textsc{pPXF} can provide much more complicated SFHs, and hence the difference between the Lick index and light-weighted \textsc{pPXF} average metallicities is not surprising. 

\subsection{Gas-phase oxygen abundances}
\label{gas_metallicities}
We use the empirical R-calibration of \citet{pilyugin2016} to derive gas-phase oxygen abundances. We note that this empirical calibration considers the absolute oxygen abundance scale given by the direct method (i.e. that based in the measurement of the electron temperature, $T_e$, of the ionized gas). It is well known that $T_e$-based empirical calibrations provide results that are systematically 1.5 - 5 times lower than those derived using calibrations based on photoionization models \citep[e.g. see][and references within]{lopez-sanchez2012}, and so care should be taken when comparing the results of this analysis to metallicities derived using photoionization models.

Emission line fluxes for this calibration  were obtained from the same 1$R_{e}$ aperture spectra as for the stellar population analysis using the SAMI data reduction pipeline and the method described in \citet{scott2018}.
Briefly, the \textsc{LZIFU} code of \citet{Ho2016} was employed, where the underlying stellar continuum was subtracted first before each emission line was fit with a Gaussian profile. All emission lines were fit simultaneously and emission line fluxes, velocities, and velocity dispersions were obtained from the Gaussian fits.

We correct the emission lines for dust as in \citet{poetrodjojo2021}, by determining $E(B-V)$, and then using $A(\lambda)$ at the emission line wavelengths to de-redden each line. 
We adopt the reddening function of \citet{fitzpatrick2007}, and assume a typical $R(V)=3.1$, along with Case B recombination. 

The 156 galaxies whose $1R_{e}$ aperture spectra were dominated by AGN emission according to the line diagnostic ratios of \citet{kewley2001} were removed from the sample. 
A S/N cut was employed such that only galaxies with S/N $>5$ in  [\ion{O}{ii}]~$\lambda\lambda$3727,29 , H$\beta$,  [\ion{O}{iii}]~$\lambda$5007,   [\ion{N}{ii}]~$\lambda\lambda$6548,84 and H$\alpha$ remained. We tested a S/N $>10$ cut and found it did not reduce the scatter in the resulting $M_{
\star}-Z_{gas}$ relation. 1660 galaxies were removed from the full SAMI sample following this S/N cut, the majority of which were passive galaxies, as can be seen from Figure~\ref{sample}.

The R-calibration of \citet{pilyugin2016} was employed to determine oxygen abundances, as replicated below:

\begin{equation}
\begin{split}
    (\rm{O/H})^{*}_{R,U} & = 8.589 + 0.022~\log(R_{3}/R_{2}) + 0.399~\log N_{2} \\
     & + (-0.137 + 0.164 \log(R_{3}/R_{2}) + 0.589 \log N_{2}) \\
     & \times \log R_{2},
\end{split}
\end{equation}
when $\log N_{2} \geqslant -0.6$
\begin{equation}
    \begin{split}
        (\rm{O/H)}^{*}_{R,L} & = 7.932 + 0.944~ \log(R_{3}/R_{2}) + 0.695 \log N_{2} \\ 
        & + (0.970 - 0.291 \log(R_{3}/R_{2}) - 0.019 \log N_{2}) \\ 
        & \times \log R_{2},
    \end{split}
\end{equation}
when $\log N_{2} < -0.6$, and where:\\ 
$R_{2} = I_{[\rm{OII}]\lambda3727 + \lambda3729}/I_{\rm{H\beta}}$, \\
$R_{3} = I_{[\rm{OIII}]\lambda4959 + \lambda5007}/I_{\rm{H}\beta}$, \\
$N_{2} = I_{[\rm{NII}]\lambda6548 + \lambda6584} / I_{\rm{H}\beta}$, \\
and $(\rm{O/H})^{*}_{R,U}$ and $(\rm{O/H})^{*}_{R,L}$ refer to the upper and lower tracks of the relation respectively, and $(\rm{O/H})^{*}_{R}\equiv 12 + \log(O/H)_{R}$.

\begin{figure}
\centering
\begin{subfigure}{0.49\textwidth}
\includegraphics[width=\textwidth]{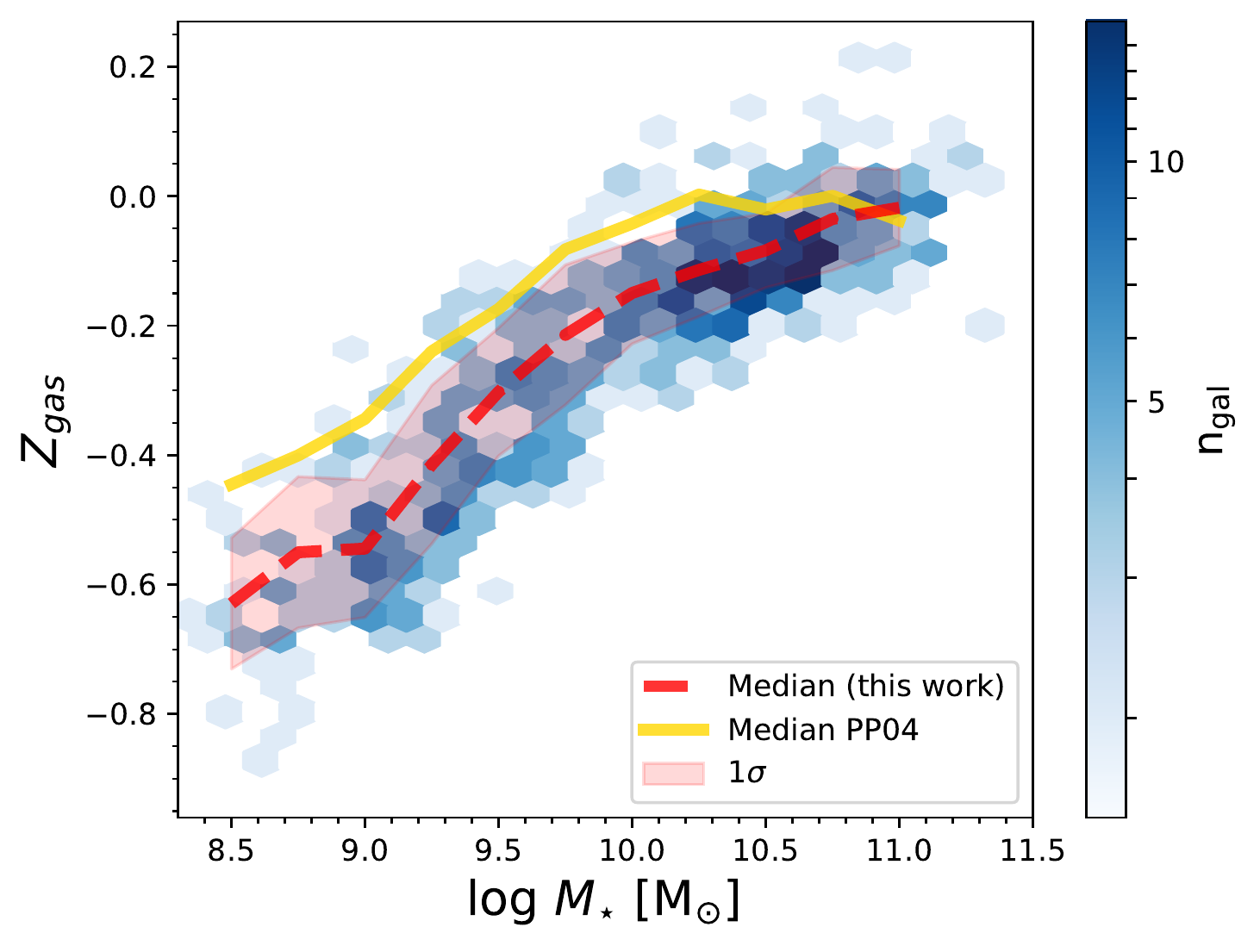}  
\end{subfigure}
\caption{The $M_{\star}-Z_{gas}$ relation for SAMI galaxies in this work derived using the empirical calibration of \citet{pilyugin2016}. Hexbins are coloured by the number of galaxies in each bin. The red dashed line denotes the median of this relation, and the red shaded region the 1$\sigma$ scatter, the average of which is 0.08 dex. For comparison, the median derived from the empirical O3N2-based calibration of \citet{pettini2004} is shown in gold. The flattening of the $M_{\star}-Z_{gas}$ relation at high $Z_{gas}$ as seen in previous works is apparent.}
\label{gas_mass_met}
\end{figure}

We define $Z_{gas}\equiv 12 + \log(\rm{O/H})_{R} - 8.76$, or $Z_{gas}\equiv \log(\rm{O/H})+3.24$, where the solar value oxygen abundance of 8.76 \citep{nieva2012} has been subtracted.
The $M_{\star}-Z_{gas}$ relation derived from the \citet{pilyugin2016} calibration is shown in Figure~\ref{gas_mass_met} for SAMI galaxies that have both $Z_{gas}$ and $Z_{\star}$ estimates, along with a comparison relation derived from the empirical O3N2 calibration of \citet{pettini2004} for the same galaxies in gold. The flattening in the $M_{\star}-Z_{gas}$ relation at high $Z_{gas}$ is apparent in Figure~\ref{gas_mass_met} as observed in previous works \citep[e.g.][]{tremonti2004a, gallazzi2005, foster2012, lara-lopez2013}.

To summarise the final sample, from the 3068 unique galaxies in the SAMI galaxy survey, 2745 have available stellar metallicities from full spectral fitting methods, 1251 have available gas-phase oxygen abundances, and 472 have both indicators \textit{and} available SFRs from GSWLC. These 472 galaxies are the sample used for the rest of this work. From Figure~\ref{sample} it can be seen that the galaxies with both stellar metallicity and gas-phase oxygen abundance measurements are representative of galaxies on the star-forming main sequence. 

\section{Results}
\label{results}
\subsection{Comparing stellar and gas-phase metallicities}

 \begin{figure}
\centering
\begin{subfigure}{0.49\textwidth}
\includegraphics[width=\textwidth]{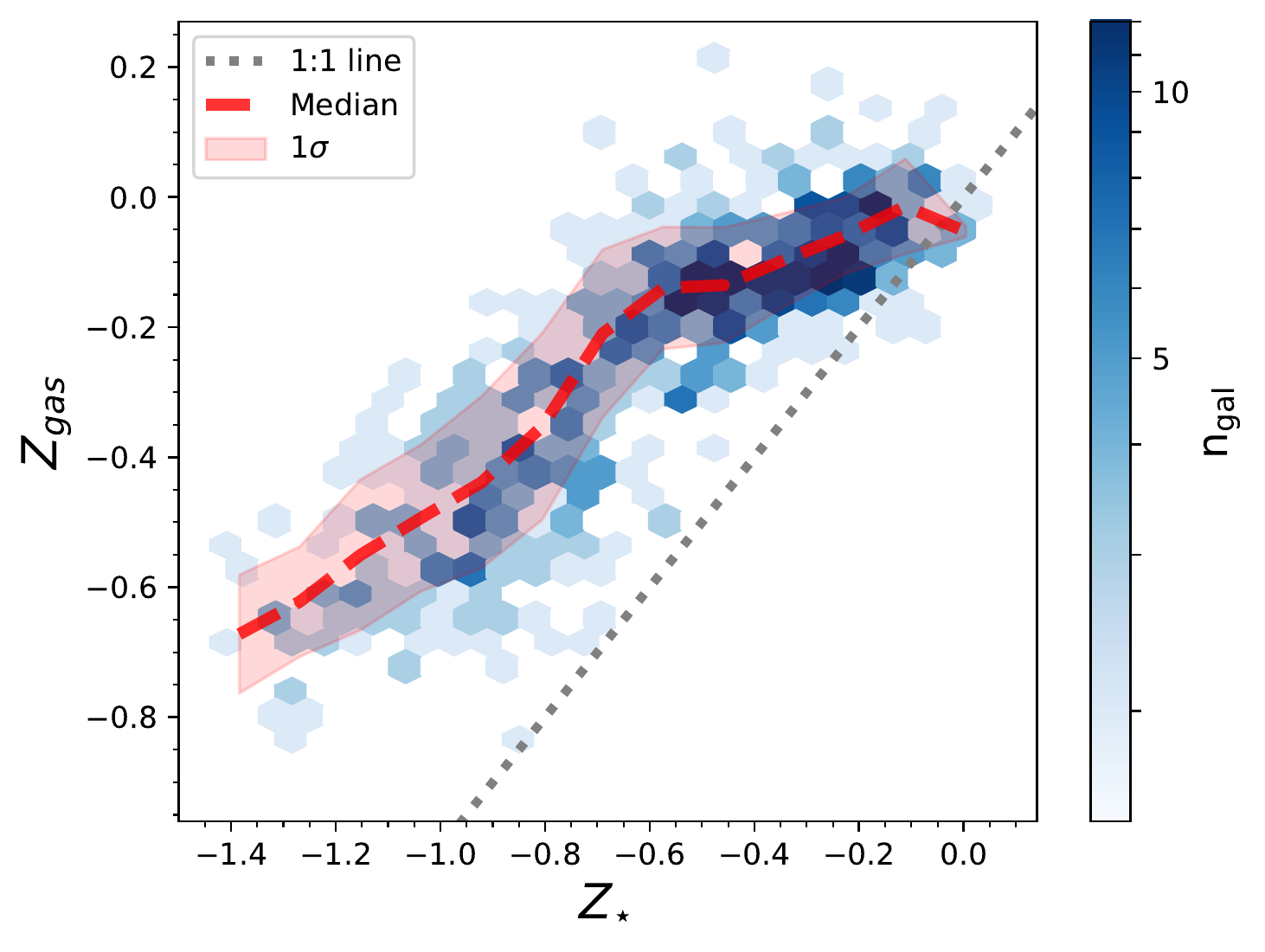}  
\end{subfigure}
\caption{Comparison of $Z_{gas}$ using the \citet{pilyugin2016} calibration to the \textsc{pPXF}-derived light-weighted $Z_{\star}$ for SAMI galaxies. Hexbins are coloured by the number of galaxies in each bin. The red dashed line denotes the running median and the shaded region is the 1$\sigma$ scatter. For all but the most stellar metal-rich galaxies, $Z_{\star}$ is always lower than $Z_{gas}$.}
\label{metmet}
\end{figure}

In Figure~\ref{metmet} we plot the \textsc{pPXF}-derived light-weighted stellar metallicity as a function of the gas-phase oxygen abundance estimates from the \citet{pilyugin2016} calibration for the SAMI galaxies with both measures available. The median of the distribution is shown as a red dashed line and the shaded red region denotes the 1$\sigma$ scatter. The grey dotted line is the 1:1 line, denoting the region where the stellar and gas-phase metallicities are equal. The relationship between $Z_{gas}$ and $Z_{\star}$ is non-linear, and there is a `knee' region above which the correlation flattens off. We expect this flattening to be due to the curvature of the $M_{\star}-Z_{gas}$ relation, and it is for the most metal-rich galaxies that the stellar metallicity approaches that of the gas-phase oxygen abundance.
The scatter in the $Z_{gas}-Z_{\star}$ relation is 0.1 dex, which is a considerable improvement on the early SDSS results of \citet{gallazzi2005}, whose 1 $\sigma$ scatter in this relation was $\sim0.3$ dex. We expect this improvement is chiefly due to the gas-phase and stellar metallicity calibrators employed. 

We define the gas-phase oxygen abundance-to-stellar metallicity ratio, $\Delta Z_{g,\star}$, as the ratio (or logarithmic difference) of the two metallicity indicators, where 
\begin{equation}
    \Delta Z_{g,\star} \equiv Z_{gas} - Z_{\star} = \log (\rm{O}/\rm{H}) + 3.24 - [\rm{M/H}].
\end{equation}

The metallicity ratio is on a logarithmic scale, such that $\Delta Z_{g,\star} = 0.5$ implies that the gas is $\sim3 \times$ as metal-rich as the stars, and $\Delta Z_{g,\star}=0$ implies that the two indicators are equal to one another.

\subsection{Correlations with $\Delta Z_{g,\star}$}

 \begin{figure*}
\centering
\begin{subfigure}{0.8\textwidth}
\includegraphics[width=\textwidth]{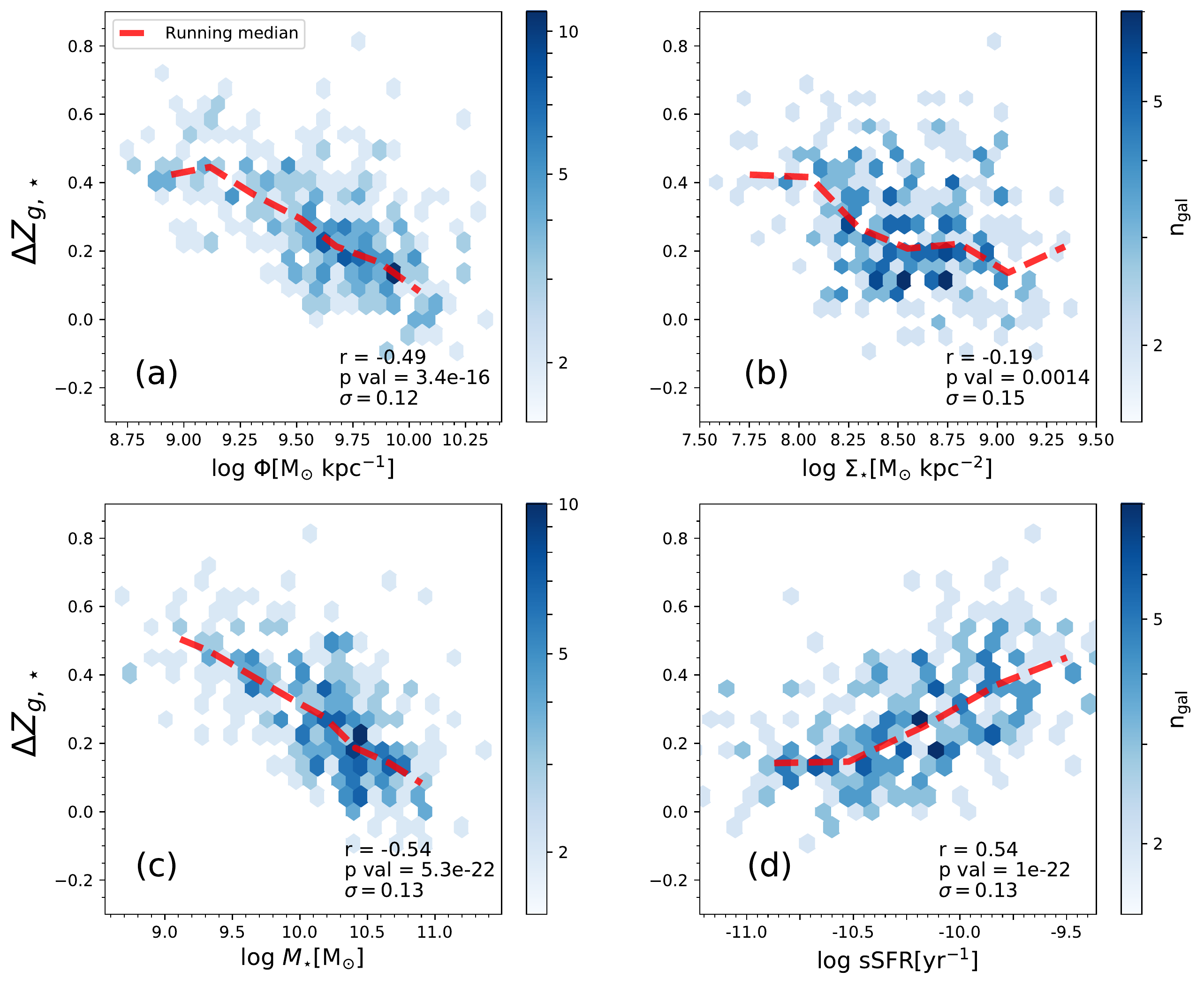}  
\end{subfigure}
\caption{The correlation between $\Delta Z_{g,\star}$ and four galaxy properties that correlate with $Z_{gas}$ and $Z_{\star}$ individually. For each panel, the Spearman correlation coefficient (r), p-value, and the vertical scatter in the correlation ($\sigma$) are shown. The red dashed line denotes the running median of the SAMI galaxies. While strong correlations and anti-correlations are seen between $\Delta Z_{g,\star}$ and $\log \Phi$, $\log \rm{M}_{\star}$, and $\log \rm{sSFR}$, only a weak anti-correlation is seen with $\log \Sigma_{\star}$.}
\label{comparisons}
\end{figure*}

With the aim of identifying why there is such a large variation in $\Delta Z_{g,\star}$ in the SAMI sample,
in Figure~\ref{comparisons} we explore $\Delta Z_{g,\star}$ as a function of some properties that are known to scale with $Z_{gas}$ and  $Z_{\star}$ individually. In panels (a) to (d), the properties chosen are: (a) $\Phi = \log(M_{\star}/R_{e})$ a proxy for gravitational potential,  (b) $\Sigma_{\star}\sim\log(M_{\star}/R_{e}^{2})$ a proxy for stellar surface mass density, (c) stellar mass $M_{\star}$, and (d) specific star formation rate ($\rm{sSFR} = \rm{SFR}/M_{\star}$).

For each panel of Figure~\ref{comparisons}, the Spearman correlation coefficient (r), p-value, and $1\sigma$ vertical scatter are displayed for each fit. Correlations are seen between $\Delta Z_{g,\star}$ and $\Phi$ (r=-0.49), $\rm{M}_{\star}$ (r=-0.54), and sSFR (r=0.54). There is only a very weak correlation between $\Delta Z_{g,\star}$ and $\Sigma_{\star}$ (r=-0.19). Interestingly, although there are different degrees of correlation, stellar mass features in all four of these quantities, and one of the strongest trends in $\Delta Z_{g,\star}$ is with stellar mass. When the additional parameter of optical size is added to obtain $\Phi$, this correlation weakens. If we increase the contribution of optical size to obtain $\Sigma_{\star}$ (which contains an $R_{e}^{2}$ term), the correlation weakens further. On the other hand, when SFR is incorporated with stellar mass to obtain sSFR, the correlation remains as tight as for with stellar mass. 

\begin{figure*}
\centering
\begin{subfigure}{0.45\textwidth}
\includegraphics[width=\textwidth]{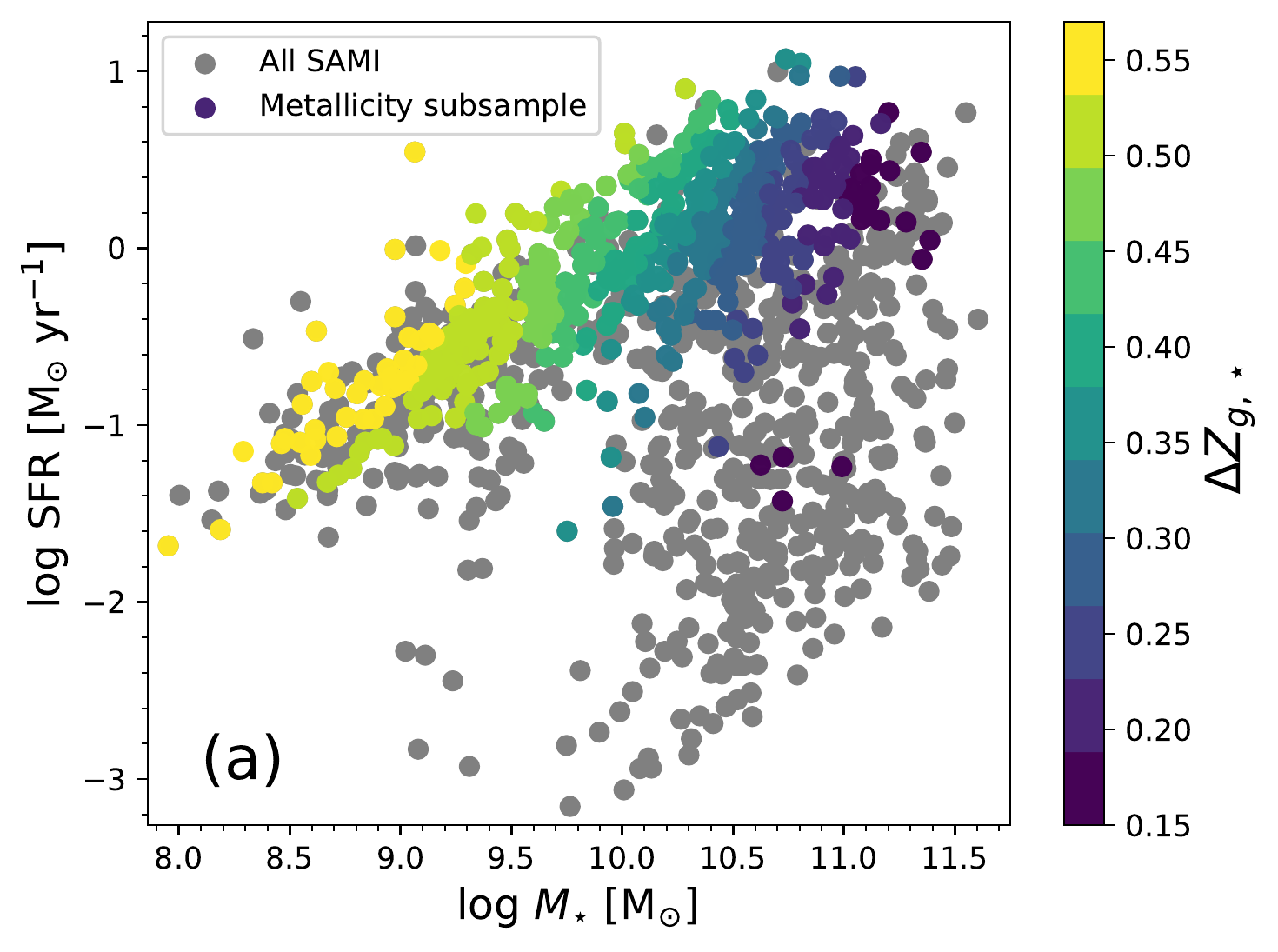}  
\end{subfigure}
~
\begin{subfigure}{0.45\textwidth}
\centering
\includegraphics[width=\textwidth]{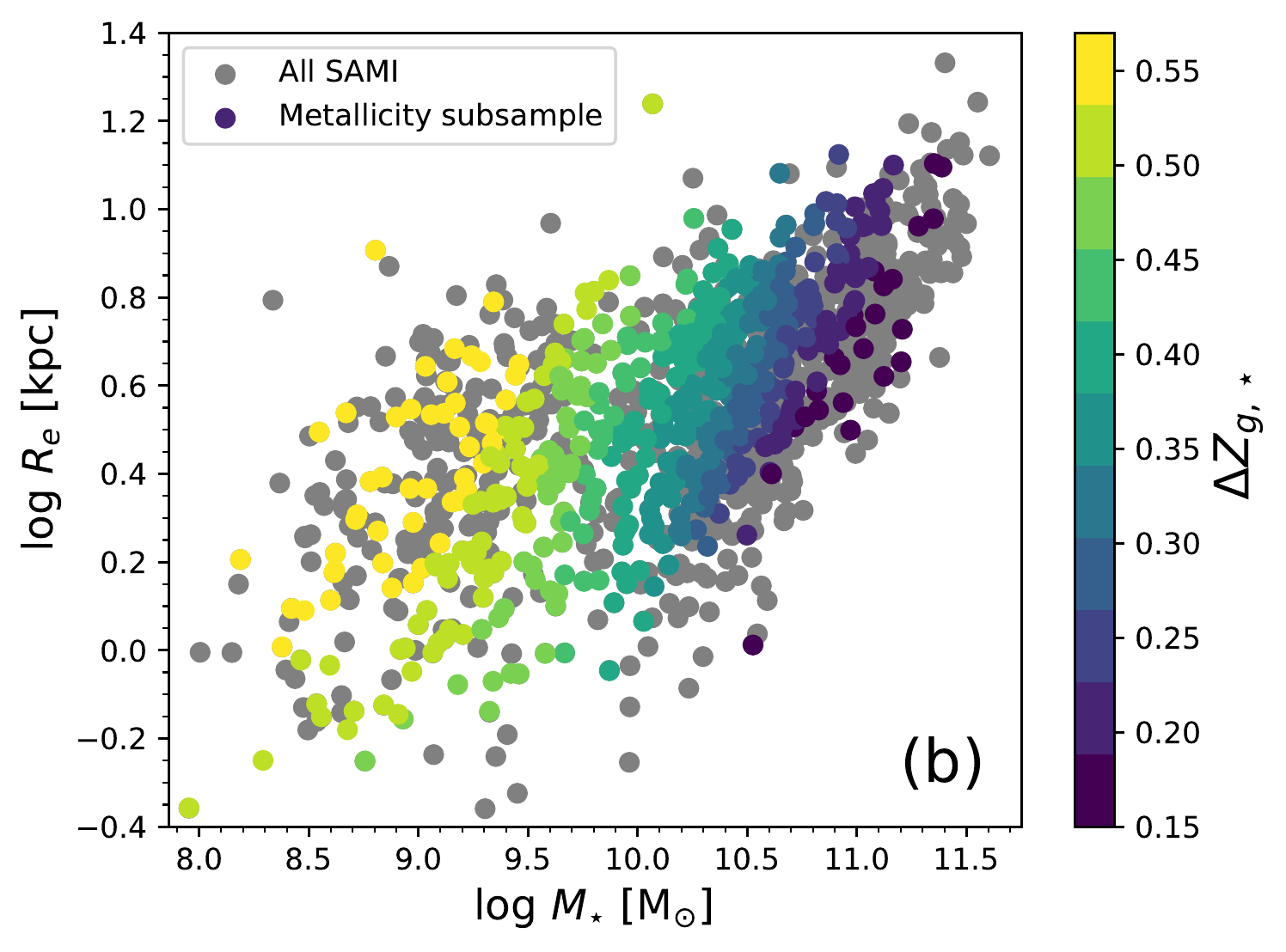}  
\end{subfigure}
\caption{Understanding the relative importance of $M_{\star}$, sSFR, and $\rm{\Phi}$ on the galaxy metallicity ratio $\Delta Z_{g,\star}$. In panel (a) we plot the star-forming main sequence, and in panel (b) the mass-size relation. The points of both plots are coloured by $\Delta Z_{g,\star}$.
Both plots are LOESS smoothed to bring out trends in each parameter. In both the star-forming main sequence and the mass-size relation, the main trend is with stellar mass, although residual trends exist in both SFR and effective radius.}
\label{results3}
\end{figure*}

We investigate links between $\Delta Z_{g,\star}$ and stellar mass, SFR, and optical effective radius further in Figure~\ref{results3}.
In panel (a) we plot star formation rate as a function of stellar mass with points coloured by $\Delta Z_{g,\star}$. 
To bring out any residual trends, we have smoothed the data using the locally weighted regression algorithm LOESS \citep[as implemented by][]{cappellari2013}.
The iso-contours of the smoothed $\Delta Z_{g,\star}$ distribution in the $M_{\star}$--SFR plane are not vertical: for a given $\rm{M}_{\star}$ there is a residual trend in $\Delta Z_{g,\star}$ with SFR. This implies that both SFR and stellar mass are responsible for the trends seen in $\Delta Z_{g,\star}$. Figure~\ref{results3} is consistent with Figure~\ref{comparisons} (d) in that galaxies with higher SFRs at fixed stellar mass also have the greatest difference between $Z_{gas}$ and $Z_{\star}$.

Panel (b) of Figure~\ref{results3} shows the mass-size relation. Again, we see residual trends in $\Delta Z_{g,\star}$ with both mass and optical effective radius ($R_{e}$). The inter-relation between stellar mass, SFR, and optical size is not surprising. We know that galaxies with smaller radii on the star-forming main sequence possess higher SFRs for a given stellar mass than for galaxies with larger sizes \citep[e.g.][]{wuyts2011}. In addition, our estimates of galaxy size are optical, so given the well-known relation between galaxy size and colour \citep[e.g.][]{vulcani2014}, any estimate of $\Phi$ or $\Sigma_{\star}$ will also encapsulate age differences. Finally, two galaxies of similar stellar mass with different gravitational potentials must by construction have different assembly histories and likely SFHs (as inferred from their sSFR in this case), a result of the differing inflow and outflow history. It is therefore nearly impossible to disentangle the one galaxy property that is responsible for the observed correlations in $\Delta Z_{g,\star}$.
For now, we can say that the strongest correlations seen are with stellar mass, followed by SFR and galaxy size.

\subsection{A step further: the correlation between $\Delta Z_{g,\star}$ and sSFR}
Given the trends in $\Delta Z_{g,\star}$ with sSFR coupled with the fact that previous literature ties steeper slopes of the $M_{\star}-Z_{\star}$ relation to the older stellar populations in galaxies \citep[e.g.][]{valeasari2009},
we expect that the trends observed in $\Delta Z_{g,\star}$ are related to the SFH of a galaxy.
 In order to understand the observed correlations between $\Delta Z_{g,\star}$ and sSFR (which can be considered a proxy for the shape of the SFH of a galaxy), we explore the metallicity history of the galaxies in the SAMI sample via full spectral fitting. We split the integrated stellar population output weights from \textsc{pPXF} into two bins, templates with ages $<1$ Gyr (though excluding the youngest template SSPs in the metallicity determination), and templates with ages $>7$ Gyr, and determine the average metallicity of both bins in the same manner as the values for the entire population. Galaxies with very little light in either the old or young template age ranges could bias metallicity results. For this reason, we perform this analysis only on galaxies that have $>10\%$ of their light in both the young and old age ranges.
 
\begin{figure}
\centering
\begin{subfigure}{0.49\textwidth}
\includegraphics[width=\textwidth]{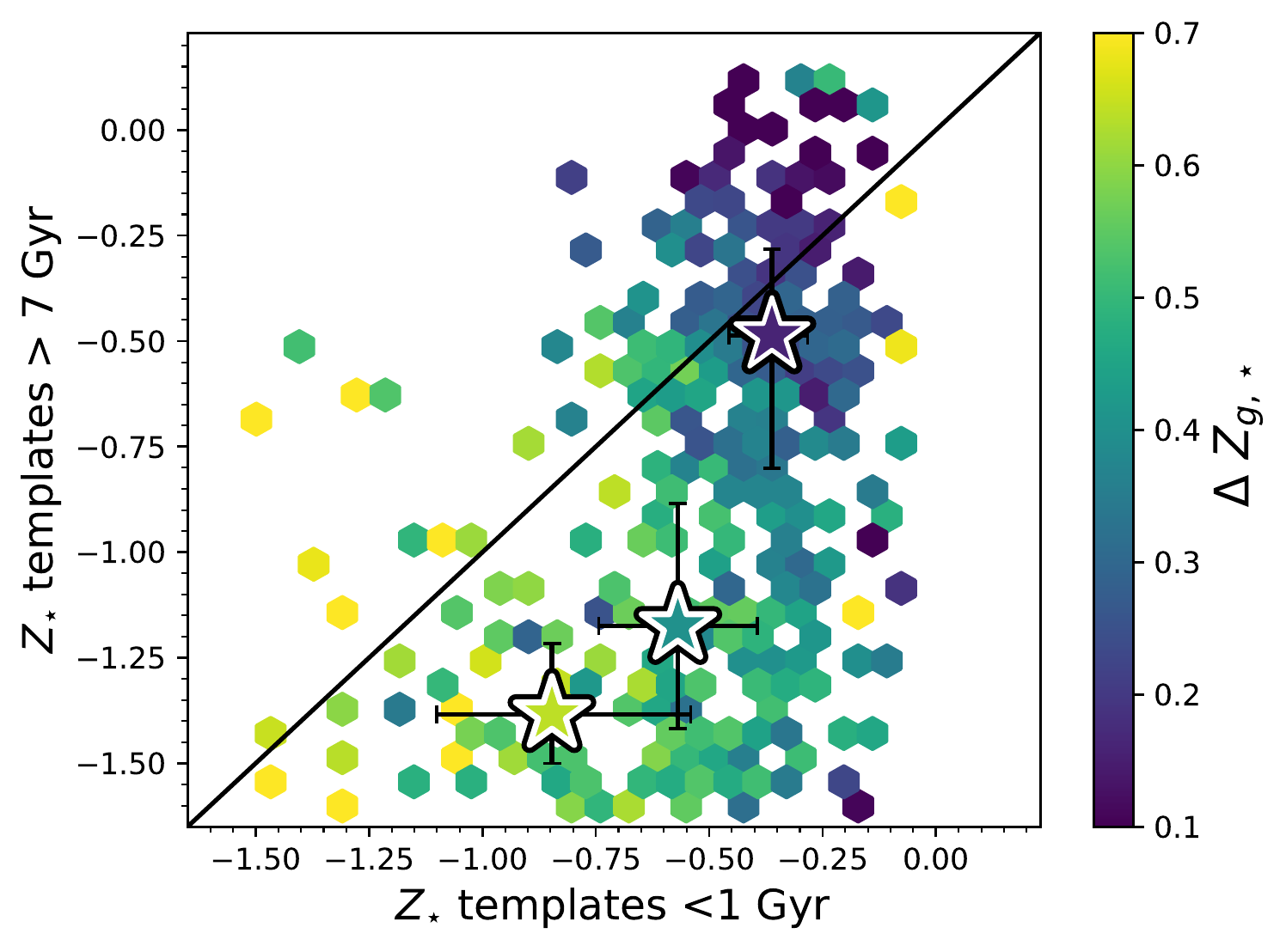}  
\end{subfigure}
\caption{Splitting the full spectral fit results into young ($<1$ Gyr) and old ($>7$ Gyr) stellar populations. The stellar metallicity $Z_{\star}$ of the $<1$ Gyr vs. $>7$ Gyr stars within the same galaxy are plotted. Points are coloured by their metallicity ratio $\Delta Z_{g,\star}$, with the large star-shaped markers representing median values in three bins of $\Delta Z_{g,\star}$. Only galaxies with $>10\%$ of light in both young and old templates are plotted.
The majority of $<1$ Gyr stars are metal rich, but the metallicity of the $>7$ Gyr stars is correlated with the metallicity ratio.}
\label{stellpopsplit}
\end{figure}

 In Figure~\ref{stellpopsplit} we show the light-weighted average stellar metallicity of the $<1$ Gyr and $>7$ Gyr stellar populations, with points coloured by $\Delta Z_{g,\star}$. 
 For clarity, we split the galaxy population into three bins of metallicity ratio:
 \begin{equation*}
 \begin{aligned}
     &\Delta Z_{g,\star} < 0.4, \\
    0.4 <  &\Delta Z_{g,\star} < 0.8, \rm{and}\\
     &\Delta Z_{g,\star} > 0.8,
     \end{aligned}
 \end{equation*}
 
 and determine the median values for the metallicity of $<1$ Gyr and $>7$ Gyr stellar populations for each bin, indicated by large star-shaped markers. 
 
 A strong vertical trend is seen whereby the $<1$ Gyr stellar populations are uniformly metal-rich, except for the greatest $\Delta Z_{g,\star}$ values where $Z_{gas}$ and $Z_{\star}$ are most different. 
 A slight `J' shape is observed in the distribution of galaxies in Figure~\ref{stellpopsplit} such that the young stellar populations of the galaxies where $\Delta Z_{g,\star} \gtrsim 0.6$ are more metal-poor than the rest of the galaxy population. From Figure~\ref{comparisons} (c) we expect these galaxies (where the stellar and gas metallicities are most different) to be low-mass dwarfs, where the galaxy hasn't had enough star formation yet to increase its overall stellar metallicity.
 
 For the overall galaxy population, the average metallicity of the $>7$ Gyr stellar populations varies according to $\Delta Z_{g,\star}$; galaxies whose average stellar metallicity is similar to their gas metallicity ($\Delta Z_{g,\star}\sim0$) possess metal-rich old stellar populations. Conversely, galaxies whose overall average stellar metallicity is much lower than their gas-phase oxygen abundance possess metal-poor old and young stellar populations. 
 
 The goal of Figure~\ref{stellpopsplit} is primarily to show that the difference between $Z_{gas}$ and $Z_{\star}$ is linked to the rapidness of the metallicity enrichment history of galaxies. Such that, galaxies with a rapid early enrichment (i.e. $Z_{\star}$ of the young stars is very similar to $Z_{\star}$ of the old stars) have $\Delta Z_{g,\star}$ closer to 0 than objects with flatter enrichment histories. 
 The strong correlation between the metallicity of the older stellar populations in a galaxy and $\Delta Z_{g,\star}$ points to the fact that it is the older stellar populations (and hence a galaxy's past star formation activity) that dictate trends seen in the metallicity ratio today. We investigate scenarios in which both high and low $\Delta Z_{g,\star}$ galaxies can occur in Section~\ref{cartoon_description}.

\section{Discussion}
\label{discussion}
\subsection{Comparing the metal content of stars and gas}
\citet{gallazzi2005} derived both stellar metallicities and gas-phase oxygen abundance estimates for 7462 fibre spectra from SDSS DR2. In almost all cases, the authors found that the stellar metallicity was lower than the gas-phase oxygen abundance, and that a large range (1$\sigma$ scatter of at least 0.3 dex) of stellar metallicities exist at fixed gas-phase oxygen abundance. 
In addition, the authors found trends in the stellar metallicity-to-oxygen abundance ratio with several other galaxy observables, including stellar mass, 4000 $\rm{\AA}$ break strength, sSFR, and gas mass fraction.
\citet{gallazzi2005} postulated that the spread of stellar metallicities at fixed gas-phase oxygen abundance was caused by the presence of gas inflows and/or gas outflows. That said, pristine gas inflows and enriched gas outflows should make the ISM more metal-poor, rendering it more similar to the stellar metallicity. In reality, both this current work and \citet{gallazzi2005} see the opposite behaviour - galaxies that are more likely to lose their gas via outflows (i.e. low $M_{\star}$ and low $\Phi$ galaxies) possess stellar and gas-phase oxygen abundances that are \textit{more different} from each other. 

\citet{lian2018} built on this previous work by aiming to create a model that explained the observed large differences in stellar metallicity and gas-phase oxygen abundances of SDSS galaxies. The authors found that either strong metal outflows or a steep IMF (both quite extreme scenarios) explained the systematically lower (mass-weighted) stellar metallicities observed when compared to gas-phase oxygen abundances.
Our results mirror those of both \citet{gallazzi2005} and \citet{lian2018}: we also report stellar metallicities at the low-mass end that are considerably lower than the gas-phase oxygen abundances in the same galaxy.

\subsection{The difference between oxygen abundances and the overall ISM metal content}

An important point rarely taken into account in previous works is that oxygen, while the most abundant heavy element of the ISM, is not its only constituent. Many other elements exist in appreciable quantities such that we cannot assume that the oxygen abundance is equivalent to the overall metallicity of the ISM (which we term here $Z_{gas,tot}$).

Traditionally, (and indeed in this paper up until now), gas-phase metallicities are inferred via oxygen abundances. Stellar metallicities are usually inferred from iron abundances, however. Directly comparing the two can be dangerous, as while oxygen is an $\alpha$ element, iron is not. $\alpha$ elements are produced in Type II supernovae on the timescale of order $\sim 10$ Myr.
In contrast, iron-peak elements are mostly produced in Type Ia supernovae, which occur on longer timescales. 

\citet{wyse1993} show that the evolution of [O/Fe] with respect to [Fe/H] is driven by the SFH of a galaxy. In particular, the `knee' of the [$\alpha$/Fe] vs. [Fe/H] relation depends on the amount of star formation early in a galaxy's life. For example, galaxies with a large initial burst of star formation will form many stars in a short amount of time, boosting $\alpha$-elements before the Type Ia supernovae can create any substantial amount of iron. Conversely, galaxies with lower sustained star formation histories will produce little oxygen before iron begins production.
These two scenarios will result in different behavior on the [$\alpha$/Fe] vs. [Fe/H] plane, and hence the comparison of iron abundances (in our case $Z_{\star}$, which can be approximated as [Fe/H]) and oxygen abundances ($Z_{gas}$) will not provide a reliable comparison of gas and stellar metal content between galaxies of differing SFHs.

 \begin{figure}
\centering
\begin{subfigure}{0.49\textwidth}
\includegraphics[width=\textwidth]{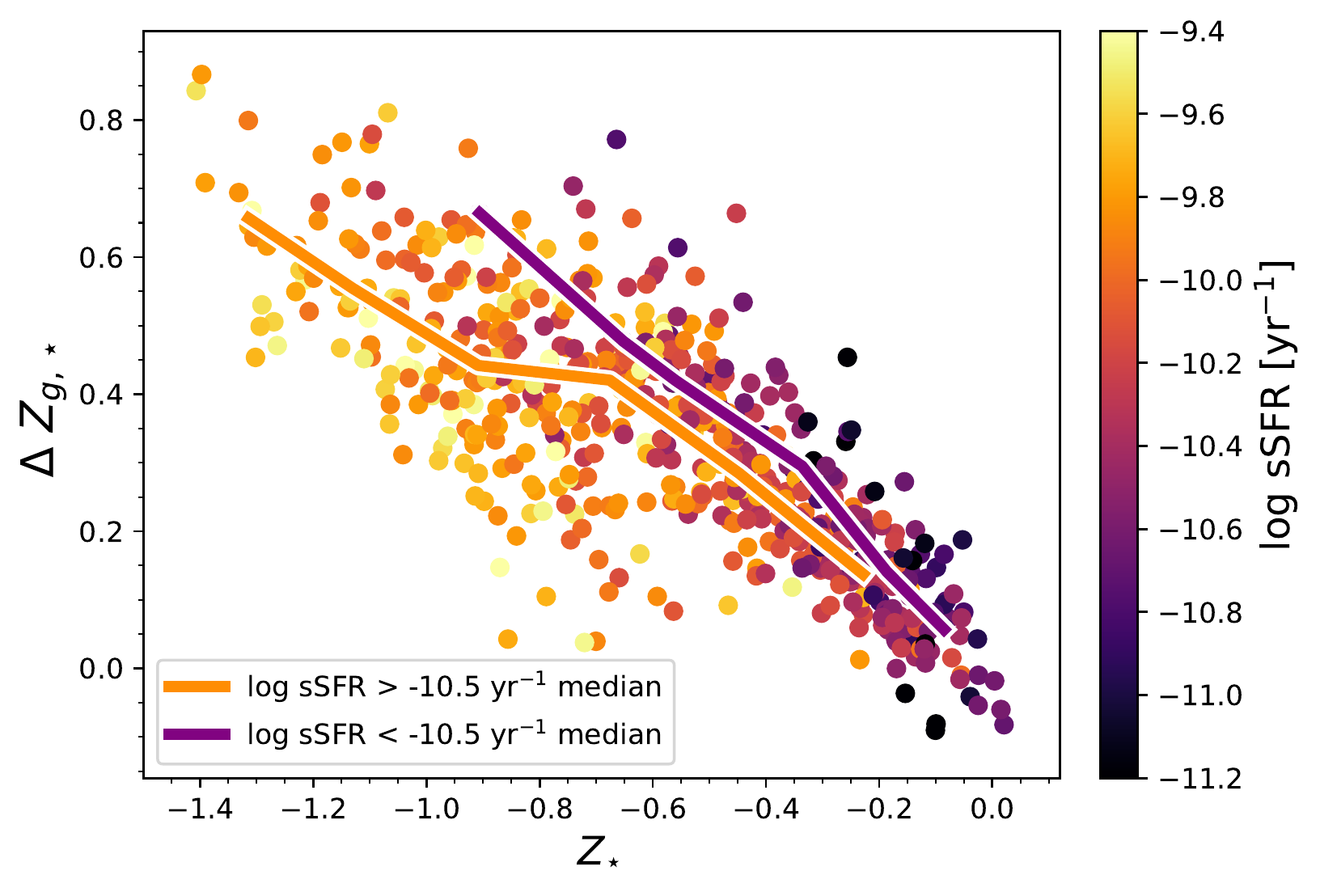}  
\end{subfigure}
\caption{The variation of the metallicity ratio $\Delta Z_{g,\star}$ as a function of stellar metallicity indicator $Z_{\star}$. Points are coloured by the sSFR of the galaxy. The medians of two bins in sSFR are shown and diverge at low $Z_{\star}$. For a given $Z{\star}$, $\Delta Z_{g,\star}$ depends on the sSFR of the galaxy. }
\label{ZZZ}
\end{figure}

Figure~\ref{ZZZ} illustrates the perils of directly comparing stellar and gas metallicities inferred from $\alpha$ and non-$\alpha$ elements to one another. At $Z_{\star}\gtrsim -0.6$, there is a tight sequence between $Z_{\star}$ and $\Delta Z_{g,\star}$. Below this value however, the scatter increases considerably, and it is difficult to tell whether a second, almost parallel sequence exists, or the relation flattens. Indeed, when the galaxy population is split into two bins of sSFR:
\begin{equation*}
 \begin{split}
    &\log \rm{sSFR} > -10.5~\rm{yr}^{-1}, \rm{and} \\
    &\log \rm{sSFR} < -10.5~\rm{yr}^{-1}.
\end{split}
\end{equation*}
The medians of these bins diverge at low $Z_{\star}$.
Clearly, at all $Z_{\star}$, the value of $\Delta Z_{g,\star}$ appears to depend on the sSFR of a galaxy. Indeed, studies such as \citet{sanchez2021} take advantage of variations in [O/Fe] as a function of [Fe/H] to describe chemical enrichment processes within galaxies.

It is clear that to directly compare the metal content of the stars and gas within galaxies, we must remove the effects shown in Figure~\ref{ZZZ} and compare metallicities based on the same element. 
While this is not possible without detailed reconstruction of the star formation and enrichment histories of every single galaxy in the sample, we can still get a feel for whether the trends shown in Figure~\ref{comparisons} are just a by-product of the different elements traced by gas and stellar metallicity measurements.

The challenge in converting oxygen abundances into a total $Z_{gas,tot}$ is that we do not know the abundance of every other element in a galaxy. 
We do however have nebular photoionisation models for the Milky Way, based on the stellar abundances of 29 nearby B stars.
\citet{nicholls2017} provide these scaling abundance patterns for \ion{H}{ii} regions for the Milky Way, Large Magellanic Cloud (LMC), and Small Magellanic Cloud (SMC) which we extrapolate for use on all SAMI galaxies. \citet{nicholls2017} derived scaling relations from the stellar spectra to describe the way in which elemental abundances scale in ionised nebulae. A result of their scaling relations are simple conversions for the base of a scaling measurement - in our case, oxygen to iron. 

The \citet{nicholls2017} work introduces a parametric enrichment factor, $\zeta$, to describe how atomic abundances scale with total abundance. This allows for conversion between scales based on different reference elements, which in our case are oxygen and iron. 
We use the online calculator\footnote{\url{https://mappings.anu.edu.au/abund/}} provided by \citet{nicholls2017} and the $\zeta$ scaling tab to convert $\log(\rm{O/H})$ to $Z_{gas,tot}$. The online tool uses the equations described in Sections 5.3 and 5.4 of \citet{nicholls2017}.

The conversion of \citet{nicholls2017} is superior to simple flat scaling relations in the literature as it takes into account how individual elements scale relative to the total nebular metallicity along with the variation of photoionisation models with nebular metal content. That said, these scaling relations rely on the chemical enrichment history of the Milky Way, LMC, and SMC only, and do not allow for other variations in abundances that could arise from different SFHs. However, the key point is that this is an improvement over a simple linear scaling which assumes a constant oxygen to metals ratio for all Z.

 \begin{figure}
\centering
\begin{subfigure}{0.49\textwidth}
\includegraphics[width=\textwidth]{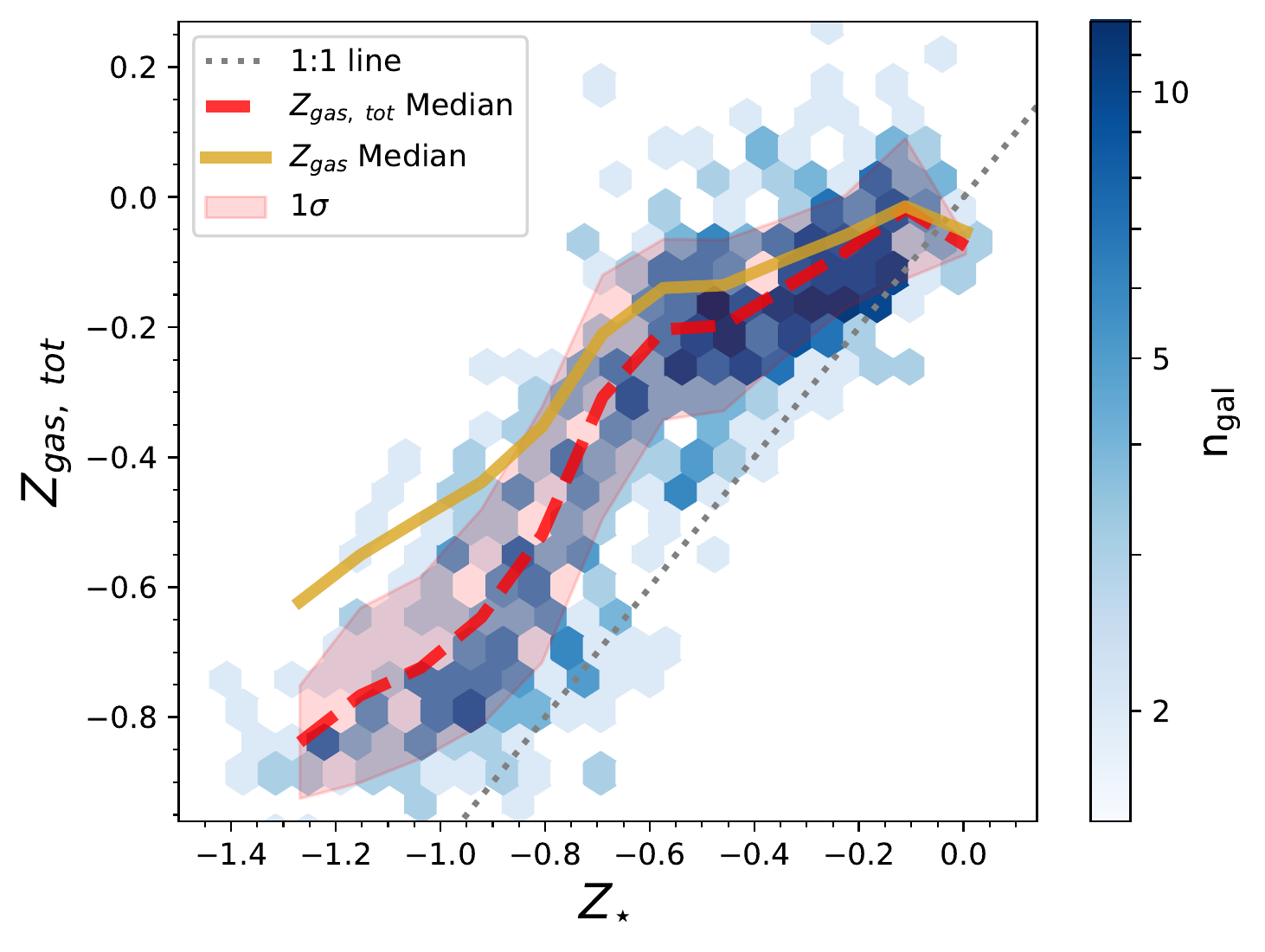}  
\end{subfigure}
\caption{Comparison of the stellar and total gas metallicity indicators for SAMI galaxies. Hexbins are coloured by the number of galaxies in each bin. The red dashed line denotes the running median. The gold line denotes the median for $Z_{gas}$ from Figure~\ref{metmet}. While $Z_{gas,tot}$ and $Z_{\star}$ are more similar to one another than $Z_{gas}$ and $Z_{\star}$ (especially at low metallicities), the stars are still almost always more metal-poor than the gas in the SAMI galaxies.}
\label{metmetZgas}
\end{figure}

 \begin{figure*}
\centering
\begin{subfigure}{0.80\textwidth}
\includegraphics[width=\textwidth]{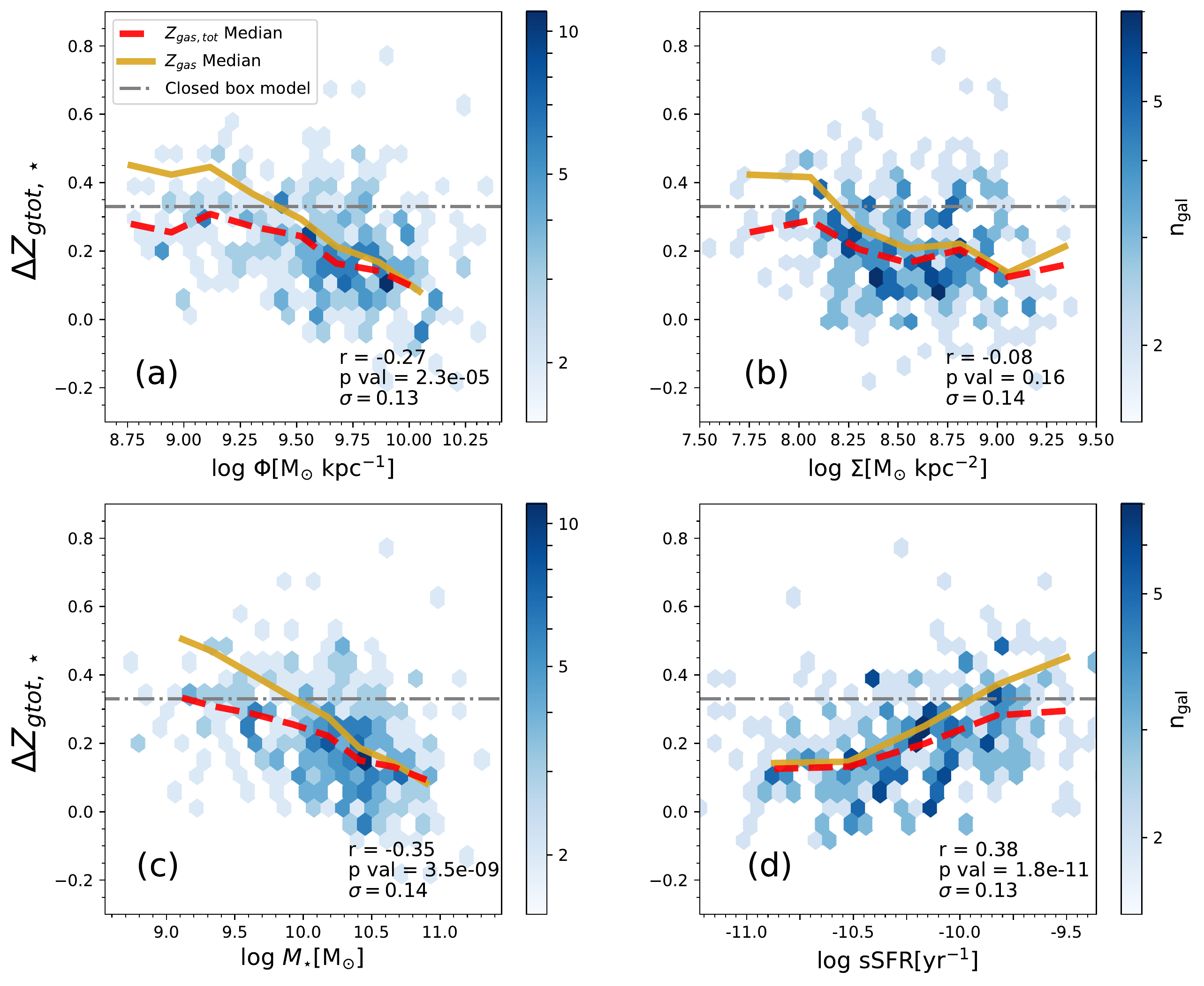}  
\end{subfigure}
\caption{Trends in $\Delta Z_{gtot,\star}$ as a function of four galaxy properties. The red dashed line is the  median of the SAMI galaxies and the gold line shows the median from Figure~\ref{comparisons}. The trends all flatten when using $Z_{gas,tot}$ instead of $Z_{gas}$, a reminder that gas and stellar abundances derived from different element bases cannot be compared between galaxies of different SFHs.}
\label{comparisons_Zg}
\end{figure*}

\subsubsection{Comparing $Z_{gas}$ and $Z_{gas,tot}$}

The results of converting from $Z_{gas}$ to $Z_{gas,tot}$ are shown in Figure~\ref{metmetZgas} with the comparison between stellar and total gas metallicities. The median trend line is shown in red, while the median line for the $Z_{gas}$ relation of Figure~\ref{metmet} is shown for comparison in gold. The relation between $Z_{gas,tot}$ and $Z_{\star}$ is closer to 1:1 than $Z_{gas}$ vs. $Z_{\star}$, and the dynamic range between $Z_{\star}$ and $Z_{gas,tot}$ is less than that of $Z_{\star}$ and $Z_{gas}$. The stars are still almost always more metal-poor than the ISM as expected. A `knee' in the $Z_{\star}-Z_{gas,tot}$ relation is still apparent and indeed enhanced with respect to the $Z_{\star}-Z_{gas}$ relation. Given the correction we are applying is just a multiplicative factor which increases with decreasing metallicity, this is not surprising. What is clear is that even with the correction, it is still true that the difference between the two metallicities is still smaller for higher metallicities.

We define $\Delta Z_{gtot,\star} = Z_{gas,tot} - Z_{\star}$, and in Figure~\ref{comparisons_Zg} we plot $\Delta Z_{gtot,\star}$ as a function of $\Phi$, $\Sigma_{\star}$, $M_{\star}$, and sSFR. The median line for $\Delta Z_{gtot,\star}$ is shown as a red dashed line, and for direct comparison to Figure~\ref{comparisons}, the median for $\Delta Z_{g,\star}$ is shown in gold. The closed box model prediction of \citet{tinsley1980} is marked by the grey dot-dashed line. 
All $\Delta Z_{gtot,\star}$ trends are flatter than $\Delta Z_{g,\star}$ trends, and this flattening primarily occurs at the lower-mass (and lower $\Phi$ and $\Sigma_{\star}$, greater sSFR) ends of the relations. 
Whilst the trends are flatter, a mild trend remains between $\Delta Z_{gtot,\star}$ and $M_{\star}$ and sSFR. 
Figures~\ref{metmetZgas} and ~\ref{comparisons_Zg} show that the average $\Delta Z_{gtot,\star}$ value is 0.35 (or the gas is $\sim2.2\times$ more metal-rich than the stars) is below the $\Delta Z_{gtot,\star}\sim0.5$ expected for a closed box model (shown as a grey dot-dashed line in Figure~\ref{comparisons_Zg}). In effect, this means that the stars are always more metal-rich (or the gas is more metal-poor) than for a model where there is no material either being deposited onto, or leaving a galaxy. Both of these options are consistent with a model in which metal-rich gas is ejected from galaxies via supernova winds, leaving enriched stars, but less enriched gas. 

Alternatively, pristine gas inflows onto a galaxy will dilute the ISM such that it is more metal-poor than expected. 
If the residual trend in $\Delta Z_{gtot,\star}$ at low stellar masses is to be believed, then it is low-mass galaxies that approach the expected value of $\Delta Z_{gtot,\star}\sim0.5$ predicted from closed box models of galaxy evolution \citep{tinsley1980}. In reality, we do not expect the low-mass galaxies to be better at holding onto their gas than their high-mass counterparts, and so it is clear that the interpretation of this ratio is more complicated than it may seem and exact values should not be blindly used to support or discard the existence of outflows/inflows.

Perhaps most importantly, the trend with mass (or potential) seems to be going in the opposite direction than expected for a simple inflow-outflow scenario, as the largest deviation from the closed box model is for high mass/high potential galaxies.
At high stellar masses it is clear that $\Delta Z_{gtot,\star}$ is lower than what is expected for a simple closed box. Thus, there must be either an infall of pristine gas or an ejection of metals into the ISM. At these masses, an infall of pristine gas would seem unlikely (and unlikely anyway at z=0) and so an outflow of metals is potentially more possible. Observational results of quenching galaxies are best modelled with contribution from a strong metal outflow at z=0 \citep[e.g.][]{lian2018, lian2018a, trussler2020}
We therefore infer that inflows and especially outflows are an important part of a galaxy's life cycle. 
 
Additionally, simple analytical models such as the gas regulator (or `bathtub') model \citep[e.g.][]{lilly2013,peng2014} predict that stellar metallicities should be $\sim0.2$ dex lower than gas metallicities at all stellar masses. It is interesting to note that this value seems most in line with what we find for galaxies of low stellar mass. 
Given these models generally include inflows and outflows but the \citet{peng2014} solution excluded variation in SFH and $Z_{gas}$, we can imply that it is not just the inflow and outflow history of a galaxy that is important in determining $\Delta Z_{g,\star}$ today, but the overall star formation and chemical evolution histories of a galaxy.

 It is clear in that converting from $Z_{gas}$ to $Z_{gas,tot}$, all trends in $\Delta Z_{gtot,\star}$ flatten. Given that \citet{wyse1993} show that the [O/Fe] abundance of a galaxy is driven by its SFH, in converting from $Z_{gas}$ to $Z_{gas,tot}$ we are removing the dependence of the gas-phase metallicity indicator on [O/Fe]. While the effects of SFH are not fully removed through this method, we can still get a feel for the effects of SFH on the trends that we observe. 
The resultant flattening of the trends in $\Delta Z_{gtot,\star}$ provides independent evidence that it is indeed the SFH of a galaxy that is driving the observed trends seen in Figures~\ref{comparisons} and~\ref{comparisons_Zg}.

\subsubsection{The link between stellar populations and SFH}
\label{cartoon_description}
\citet{fernandes2007} studied the evolution of stellar metallicity in galaxies via the derivation of SFHs from full spectral fitting.
The authors found that the star formation history of a galaxy is closely linked to its current stellar chemical abundances (and therefore the ISM metallicity). Galaxies whose ISM is metal-poor have relatively sustained star formation histories: they form stars slowly over a long period of time. In contrast, galaxies that are metal-rich assembled most of their mass at early epochs in a large burst \citep[e.g.][]{thomas2005, mcdermid2015}. 

Our results support the \citet{fernandes2007} result while adding the additional constraint of the current-day oxygen abundances. Figure~\ref{stellpopsplit} shows that the star formation history must indeed impact $\Delta Z_{g,\star}$ measured at $z=0$: we see a large difference between $Z_{gas}$ and $Z_{\star}$ at low $Z_{\star}$ for the older stellar populations and a small difference at higher $Z_{\star}$ within the older stellar population.

\begin{figure}
\centering
\begin{subfigure}{0.49\textwidth}
\includegraphics[width=\textwidth]{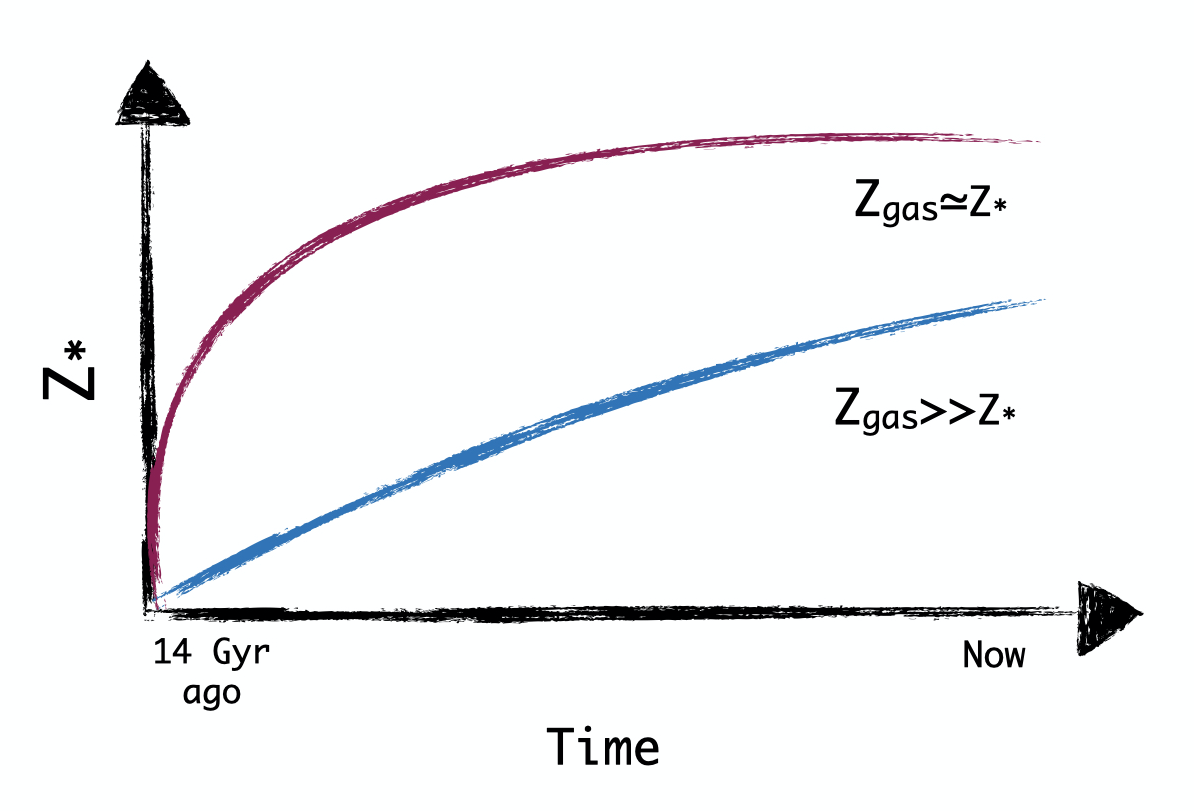}  
\caption{}
\end{subfigure}
~
\begin{subfigure}{0.49\textwidth}
\includegraphics[width=\textwidth]{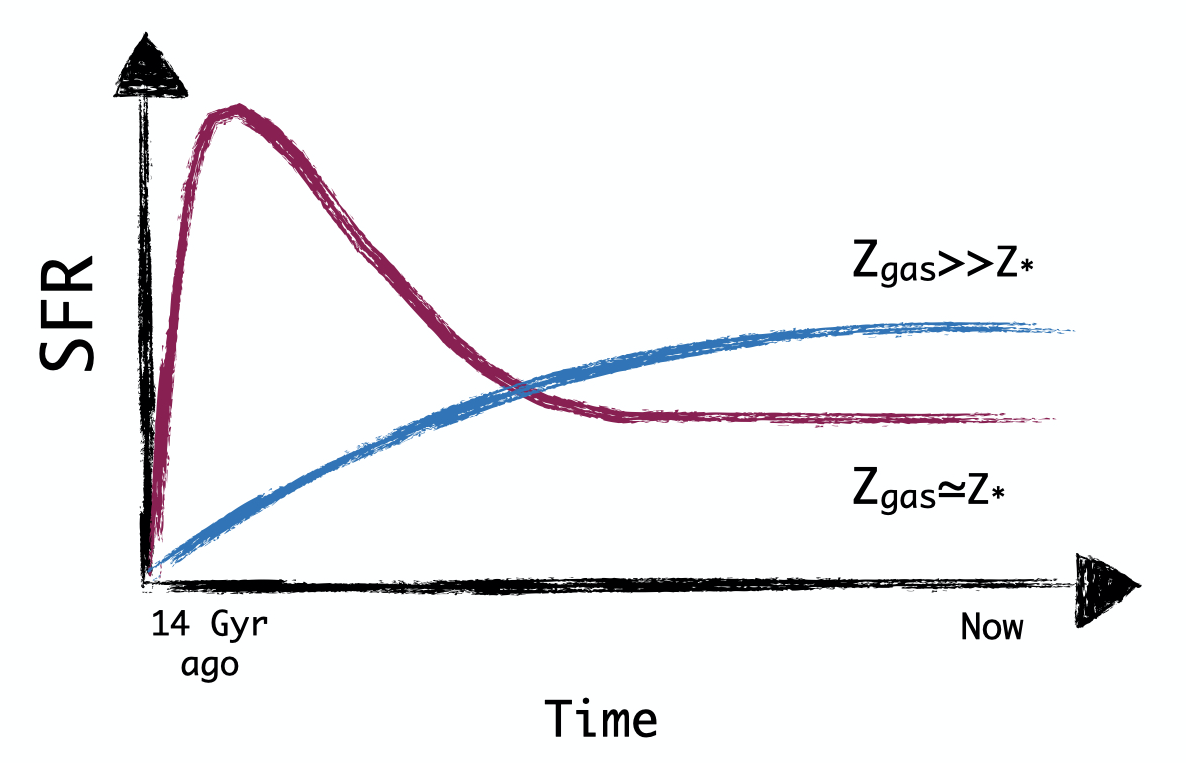}  
\caption{}
\end{subfigure}
\caption{A cartoon representation of possible chemical evolution and star formation histories for two galaxies. The red line represents a galaxy with $\Delta Z_{g,\star}\sim1$ that enriched early with an early burst of star formation. The blue line represents a galaxy where $\Delta Z_{g,\star}<<1$ at $z=0$ and had a more sustained metal enrichment with low-level star formation. Panel (a) shows two possible metal enrichment pathways, and (b) the corresponding SFHs.}
\label{cartoon}
\end{figure}

In Figure~\ref{cartoon} we provide a cartoon illustration of possible evolutionary scenarios for two galaxies at $z=0$. The red line represents a galaxy which has $\Delta Z_{g,\star}\sim0$, while the blue line represents a galaxy whose stars are much more metal-poor than its ISM ($\Delta Z_{g,\star}>0$). Panel (a) of Figure~\ref{cartoon} shows the metallicity evolution of the two galaxies as a function of time and panel (b) shows the star formation histories required to produce the observed metallicity evolution. In panel (a), the red line represents a galaxy that enriched early. Given the link between star formation and chemical enrichment, we can assume that this galaxy underwent a burst of star formation early (as shown in Figure~\ref{cartoon} (b). Assuming the burst is long enough for a number of generations of stars to form and re-enrich the ISM such that the stars at the end of the burst are formed metal-rich, the average stellar metallicity of the old stellar populations will be metal-rich, similar to both its young stellar populations and gas-phase abundance at current times. A short burst of early star formation is expected for high-mass galaxies, and we note in Figure~\ref{comparisons} (c) that it is high-mass galaxies that on average have $\Delta Z_{g,\star}\sim0$.

The blue line of Figure~\ref{cartoon} (a) represents a galaxy that has undergone a much more sustained chemical enrichment. We expect the star formation history of this galaxy to be much more prolonged, without the initial burst of star formation. As a result, the old stellar populations are on average metal-poor, and while the average $<1$ Gyr stellar populations are more metal-rich, they are not as enriched as the gas at current times. For this reason, $\Delta Z_{g,\star}>0$. Low-mass galaxies are known to have sustained bouts of star formation throughout their lifetimes, and we note in Figure~\ref{comparisons} (c) that it is low-mass galaxies that on average have $\Delta Z_{g,\star}>0$. A galaxy's instantaneous SFR is strongly linked to the gas availability at a given time. It therefore follows that while the galaxy represented by the red line must have had substantial amounts of gas available to it at early times, it is the galaxy represented by the blue line that must have a greater amount of gas available now.

It is the shape of the SFH, which in turn is linked to the stellar mass of a galaxy, that determines the chemical evolution history of a galaxy, and hence the differences seen between stellar metallicity and gas-phase oxygen abundance.

\section{Summary \& Conclusions}
\label{conclusion}
We compare  two commonly derived metallicity indicators, the stellar metallicity ($Z_{\star}$) and the gas-phase oxygen abundance ($Z_{gas}$) to one another for 472 galaxies in the SAMI galaxy survey. 
We define the ratio of gas-phase to stellar abundances, $\Delta Z_{g,\star}$, and find that $\Delta Z_{g,\star}$ correlates with $M_{\star}$, sSFR (a proxy for galaxy SFH), and $\Phi$. We see a weak correlation with $\Sigma_{\star}$. 

We split full spectral fitting results into young ($<1$ Gyr) and old ($>7$ Gyr) stellar ages and determine the metallicity of these populations. 
While younger stars are uniformly metal-rich, it is the metallicity of the older stars that dictates $\Delta Z_{g,\star}$.
We infer a scenario whereby $\Delta Z_{g,\star}$ is strongly influenced by a galaxy's star formation history: a galaxy with an early burst of star formation will build up its metals early, enriching its older stellar populations. As a result of the early enrichment, $Z_{\star}~\sim~Z_{gas}$. Conversely, galaxies with a delayed star formation history that reach their peak SFR at later times will take longer to build up their metals. Older stellar populations will be less metal-enriched, and $Z_{\star}~<<~Z_{gas}$.

Metallicities derived via different elemental abundance indicators should not be blindly compared when the elements are produced by processes occurring on different timescales. In particular, the ratio of stellar metallicities derived via iron abundances and gas-phase metallicities derived via oxygen abundances will differ depending on the SFH of a galaxy. Indeed, we show that the $\Delta Z_{g,\star}$ of galaxies at fixed $Z_{\star}$ depends on the sSFR of a galaxy. 

We attempt to convert oxygen abundances to $Z_{gas,tot}$ using the relation of \citet{nicholls2017}. 
The resulting ratio of $\Delta Z_{gtot,\star}$ is flatter than that with $\Delta Z_{g,\star}$, and greater than that expected for a closed box model. 
We conclude that this is direct evidence for the importance on inflows and outflows within galaxies.

Our results imply that the ratio of $\Delta Z_{g,\star}$ depends on the chemical enrichment history of a galaxy, of which sSFR and stellar mass are the best observational proxies.
While inflows and outflows at current times may help regulate the fraction of metals in the ISM, it is a galaxy's star formation history that determines the ratio of stellar metallicity and gas-phase oxygen abundances observed today. Our work confirms that $Z_{gas}$ and $Z_{\star}$ provide us with very different information on galaxy evolution and that they cannot be blindly compared to one another, nor can one be used to make assumptions about the other. 

\appendix 
\section{Mass-weighted stellar metallicity measures}
\label{appendixa}
\begin{figure}
\centering
\begin{subfigure}{0.49\textwidth}
\includegraphics[width=\textwidth]{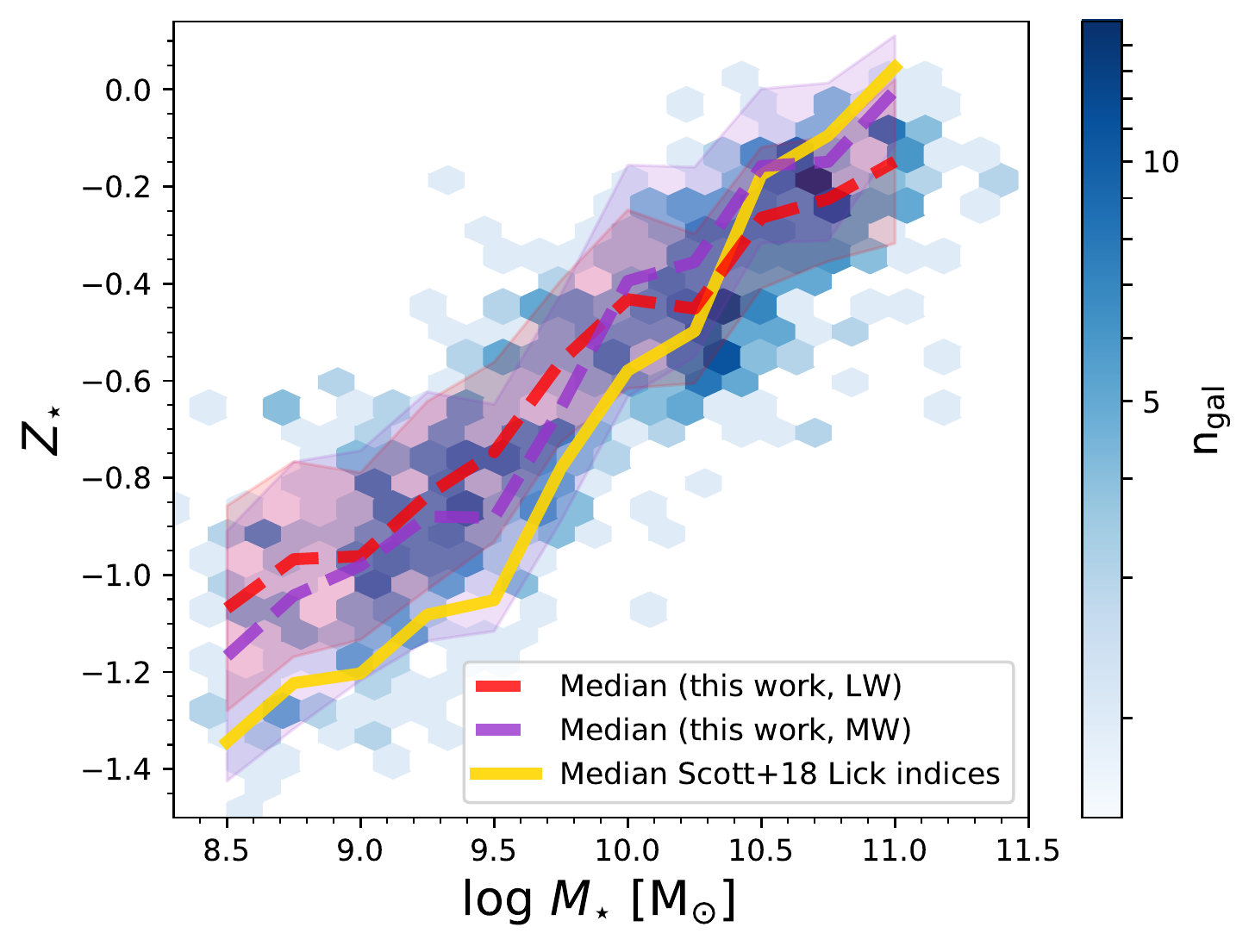}  
\end{subfigure}
\caption{Same as Figure~\ref{stell_mass_met}, but showing the comparison between \textsc{pPXF} metallicity measures derived via light-weighting (red median line, as in Figure~\ref{stell_mass_met}), mass-weighting (hexbins and purple median line), and those derived via Lick indices from \citet{scott2018}. }
\label{MZR_mw_lw}
\end{figure}

\begin{figure*}
\centering
\begin{subfigure}{0.80\textwidth}
\includegraphics[width=\textwidth]{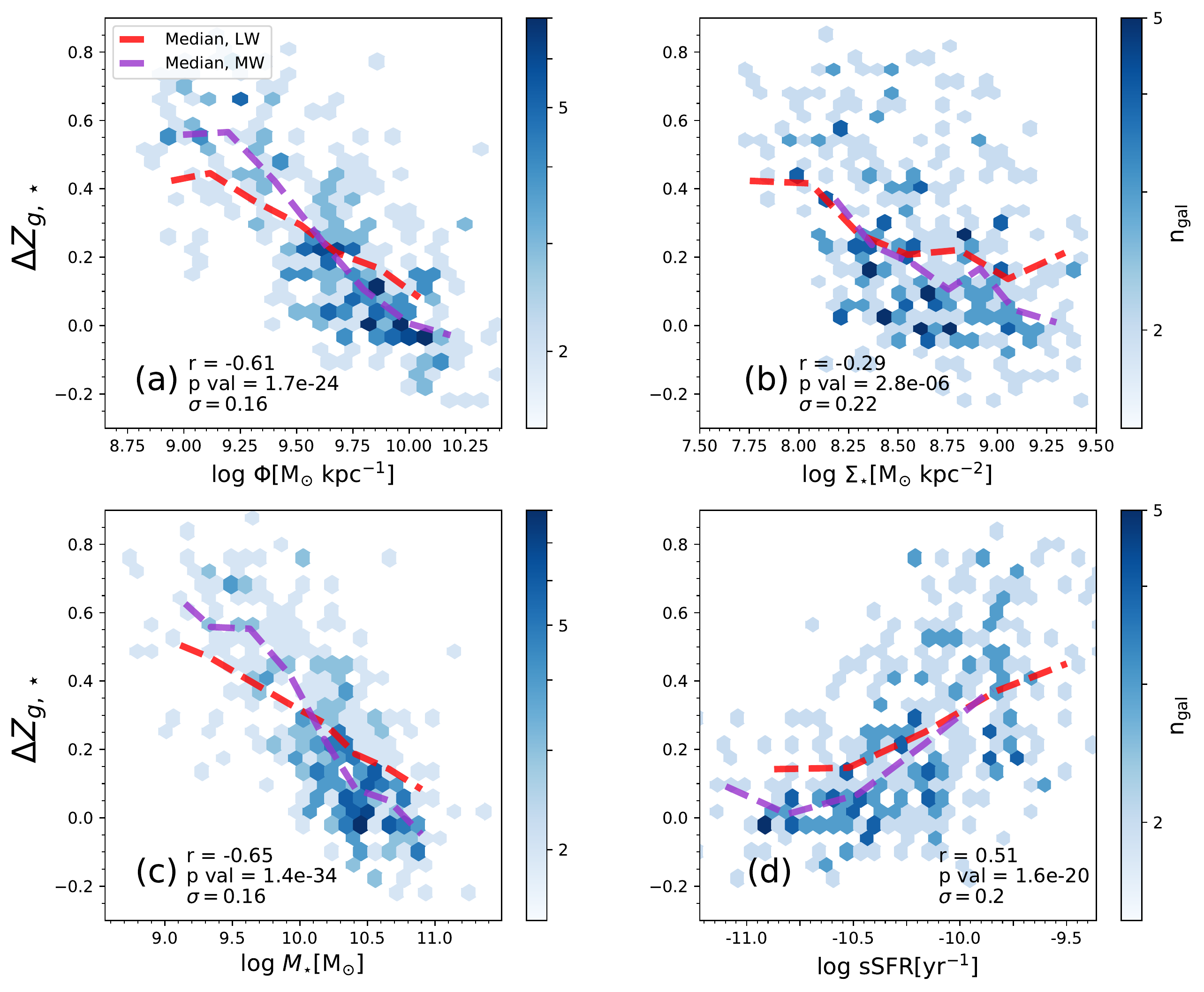}  
\end{subfigure}
\caption{Same as Figure~\ref{comparisons}, but with $\Delta Z_{g,\star}$ derived from \textbf{mass-weighted} metallicity measurements. The purple dashed line denotes the median mass-weighted values for a given property, and the red line the light-weighted values as in Fig~\ref{comparisons}. Blue hexbins are coloured by the number of galaxies per bin and are for the mass-weighted measurements.}
\label{comparisons_mw}
\end{figure*}

Light (or luminosity)-weighted stellar metallicity measures derived via full spectral fitting methods are biased towards the younger (and brighter) stellar populations in a galaxy, enabling comparison to observational quantities such as Lick indices and emission line diagnostics. Mass-weighted measures instead trace the distribution of stellar mass within a galaxy and therefore better trace the metal content of old stars within a galaxy. While useful for comparison with theoretical models, there are more assumptions involved in their derivation. For comparative purposes, we present Figures~\ref{MZR_mw_lw} and ~\ref{comparisons_mw}, which may be directly compared to the light-weighted metallicity measures as shown in Figures~\ref{stell_mass_met} and ~\ref{comparisons}. 

Importantly, from Figure~\ref{MZR_mw_lw} we see that there is little difference in the median MZR between mass-weighted and light-weighted measures: mass-weighted metallicities are $\sim0.1$ dex higher than light-weighted for galaxies of $\log M_{\star}[\rm{M}_{\odot}]>10.5$ and the reverse is true for $\log M_{\star}[\rm{M}_{\odot}]<9.5$
, meaning that the main results of this paper are not affected by choice of metallicity determination. The behaviour of the median mass-weighted and light-weighted metallicity values is interesting, but a full analysis of which is reserved for future work.

If mass-weighted metallicities are used to calculate $\Delta Z_{g,\star}$ as in Figure~\ref{comparisons_mw}, we see that all trends with $\Phi$, $\Sigma_{\star}$, $M_{\star}$, and sSFR steepen, though scatter increases by between 0.03-0.07 dex. When light-weighted metallicities were considered, we determined that mass has the strongest correlation with $\Delta Z_{g,\star}$. This is still the case when mass-weighted metallicities are used, though $\Phi$ now exhibits a stronger trend than sSFR. While both properties are doubtless important in determining the ratio of stellar to gas-phase metallicity within a galaxy, there may be a residual trend between mass- and light-weighted metallicity measures as a function of the galaxy's current SFR which may be affecting trends seen in panel (d).

\section*{Acknowledgements}
The authors wish to thank the anonymous referee for their insightful comments that have improved the impact of this work.
The authors also wish to thank Jianhui Lian for helpful conversations about his previous work. 

The SAMI Galaxy Survey is based on observations made at the Anglo-Australian Telescope. The Sydney-AAO Multi-object Integral field spectrograph (SAMI) was developed jointly by the University of Sydney and the Australian Astronomical Observatory. The SAMI input catalogue is based on data taken from the Sloan Digital Sky Survey, the GAMA Survey and the VST ATLAS Survey. The SAMI Galaxy Survey is supported by the Australian Research Council Centre of Excellence for All Sky Astrophysics in 3 Dimensions (ASTRO 3D), through project number CE170100013, the Australian Research Council Centre of Excellence for All-sky Astrophysics (CAASTRO), through project number CE110001020, and other participating institutions. The SAMI Galaxy Survey website is http://sami-survey.org/. 
LC is the recipient of an Australian Research Council Future Fellowship (FT180100066) funded by the Australian Government.
SB acknowledges funding support from the Australian Research Council through a Future Fellowship (FT140101166).
JJB acknowledges support of an Australian Research Council Future Fellowship (FT180100231).
FDE acknowledges funding through the ERC Advanced grant 695671 ``QUENCH'', the H2020 ERC Consolidator Grant 683184 and support by the Science and Technology Facilities Council (STFC).
JvdS acknowledges support of an Australian Research Council Discovery Early Career Research Award (project number DE200100461) funded by the Australian Government.
JBH is supported by an ARC Laureate Fellowship FL140100278. The SAMI instrument was funded by Bland-Hawthorn's former Federation Fellowship FF0776384, an ARC LIEF grant LE130100198 (PI Bland-Hawthorn) and funding from the Anglo-Australian Observatory.
M.S.O. acknowledges the funding support from the Australian Research Council through a Future Fellowship (FT140100255).

\section*{Data Availability}
The SAMI Galaxy Survey data release 3 (DR3) is available online at \url{https://datacentral.org.au/}. The additional data products derived for this work including stellar and gas-phase metallicity estimates can be provided on request.

    \bibliographystyle{mnras}
  \bibliography{Metals.bib}
\end{document}